%% file: df-skinter.tex
\documentclass[a4paper]{extarticle}
\usepackage[english]{babel}
\usepackage[pdftex]{graphicx}
\usepackage[pdftex]{color}
\usepackage{pstricks}
\usepackage{units}
\usepackage[utf8x]{inputenc}
\usepackage{amsmath}
\usepackage{amsfonts}
\usepackage{amscd}
\usepackage{amssymb}
\usepackage{mathrsfs}
\usepackage{amsthm}
\usepackage{url}
\usepackage{floatflt}
\usepackage{epic}
\usepackage{rotating}

\newcommand{\de}{\mathrm{d}}
\newcommand{\e}{\mathrm{e}}

\newcommand{\SU}{\mathrm{SU}}
\newcommand{\SO}{\mathrm{SO}}
\newcommand{\Oh}{\mathrm{O}}

\newcommand{\Tr}{\mathrm{Tr}}
\newcommand{\IR}{\mathbb{R}}

\newcommand{\IZ}{\mathbb{Z}}

\usepackage{lscape}
\usepackage[top=3.2cm,bottom=2.6cm,left=4.0cm,right=4.0cm]{geometry}
\usepackage{afterpage}
\definecolor{DarkRed}{rgb}{0.7,0,0}
\definecolor{DarkBlue}{rgb}{0,0,0.7}
\begin{document}
\global\emergencystretch = .9\hsize
\DeclareGraphicsRule{.pdftex}{pdf}{*}{}
 
\pagestyle{plain}
\title{\vskip -70pt
\begin{flushright}
{\normalsize DAMTP-2011-105} \\
\end{flushright}
\vskip 50pt
{\bf \Large Interactions of $B=4$ Skyrmions}
 \vskip 30pt}
\author{Dankrad T.J. Feist\thanks{D.Feist@damtp.cam.ac.uk}\\ \\
{\sl Department of Applied Mathematics and Theoretical Physics,}\\
{\sl University of Cambridge,}\\
{\sl Wilberforce Road, Cambridge CB3 0WA, England.}\\
}
\vskip 20pt
\date{December 2011}
\maketitle
\vskip 20pt

\begin{abstract}
It is known that the interactions of single Skyrmions are asymptotically described by a Yukawa dipole potential. Less is known about the interactions of solutions of the Skyrme model with higher baryon number. In this paper, it is shown that Yukawa multipole theory can be more generally applied to Skyrmion interactions, and in particular to the long-range dominant interactions of the $B=4$ solution of the Skyrme model, which models the $\alpha$-particle. A method that gives the quadrupole nature of the interaction a more intuitive meaning in the pion field colour picture is demonstrated. Numerical methods are employed to find the precise strength of quadrupole and octupole interactions. The results are applied to the $B=8$ and $B=12$ solutions and to the Skyrme crystal.
\end{abstract}

\vskip 80pt

\newpage

\section{Introduction}

50 years ago, T.H.R. Skyrme suggested a nonlinear scalar field theory as a model for nucleons and nuclear interactions \cite{skyrme1961}. While quantum chromodynamics (QCD) is currently accepted as the fundamental theory of nuclear interactions, it is nearly impossible to work with to model actual nuclei. Therefore, except for very few results derived from first principles (e.g. in lattice field theory), to this date, phenomenological theories are mainly used in nuclear physics. The Skyrme model is a field theory intermediate betweeen QCD and the point particle quantum mechanics of nucleons.

After the Skyrme model was first applied by Adkins, Nappi and Witten \cite{adkins1983} to derive the properties of single nucleons and delta resonances, it has been used to derive the properties of several nuclei. The Skyrme model with massive pions has in particular been applied to many nuclei with a baryon number divisible by four, in which the nuclei are thought to be composed of a number of interacting $\alpha$-particles \cite{battye2006}. The classical $B=4$ Skyrmion solution corresponding to the $\alpha$-particle has a cubic shape and is strongly bound. In the case of $B=8,12$ and $32$, the $B=4$ subunits have been found to be in face-to-face contact, to which the methods in this paper can be directly applied to find interaction energies.

\subsection{The Skyrme model}

The Skyrme model \cite{skyrme1961} is a three-dimensional non-linear sigma model. It includes an extra term which is necessary for having solitons according to Derrick's theorem. It is easiest to write by taking the Skyrme field $U(x)$ to be an $\SU(2)$-valued field. The action then takes the form
\begin{equation} \label{eq:skyrme}
S = \frac{1}{12 \pi^2}\int \de^4 x \left( - \frac{1}{2} \Tr(L_\mu L^\mu) + \frac{1}{16} \Tr([L_\mu,L_\nu][L^\mu,L^\nu]) + m^2 \Tr(U-1) \right)
\end{equation}
where $L_\mu=U^\dagger (\partial_\mu U)$ and the metric signature used throughout is $(+,-,-,-)$. This is the rescaled form of the theory: classical energy units are $12 \pi^2 F_\pi/(4e)$ and length units $2/(e F_\pi)$ where $F_\pi$ is the pion decay constant and $e$ a parameter of the Skyrme model. The constant $m$ is the mass of the pion in length units $2/(e F_\pi)$. The $m$-term forces $U\rightarrow 1$ as $x$ goes to spatial infinity; in the massless case ($m=0$) the same is used as a boundary condition. Note that the action in equation (\ref{eq:skyrme}) can also be written in terms of the right-current $R_\mu=(\partial_\mu U) U^\dagger$ (it has the same form replacing $L$ with $R$). The field equation can be derived \cite{manton2004} to be
\begin{equation}\label{eq:field}
\Tr \ \tau_k \left(\partial_\mu \left(L^\mu + \frac{1}{4}[L^\nu,[L_\nu,L^\mu]] \right)+ m^2 U \right)= 0 \text{,}\quad k=1,2,3 \text{ ,}
\end{equation}
which is also valid for the right current $R^\mu$. Here $\tau_k$ are the Pauli matrices. In addition to the spatial degrees of symmetry, it is also invariant under the isospin rotation $U(x) \rightarrow AU(x)A^{-1}$ where $A$ is an $\SU(2)$ matrix.

This model has been suggested by Skyrme \cite{skyrme1961} as a theory of pions and baryons (here: the proton, neutron and delta resonances which can be made from just up and down quarks). It is now believed to be a low-energy effective theory for quantum chromodynamics, although a formal derivation has not yet been achieved.

$\SU(2)$ is topologically the same as $S^3$, as can be seen by writing $U=\sigma + i \vec \pi \cdot \vec \tau$ with the restriction $\sigma^2+\vec\pi^2 = 1$, so that $(\sigma, \vec \pi)$ are coordinates on $S^3$. Pions are found in this model as the perturbative waves of the form $U \approx 1+i \vec \pi \cdot \vec \tau$, and their mass is given by $m$. Baryons occur as topological solitons. Considering $U$ at any instant of time, it is a map $U:\IR^3 \rightarrow \SU(2)\cong S^3$. As $U(\infty)=1$ has a definite value, $\IR^3$ can be compactified using the one-point compactification $\IR^3 \cup \{\infty\}\cong S^3$. The field $U$ then gives rise to a map $S^3 \rightarrow S^3$ that has an associated degree $B\in\IZ$. In fact there is also a (topologically) conserved current
\begin{equation}
B^\mu = - \frac{1}{24 \pi^2} \epsilon^{\mu \nu \sigma \tau} \Tr(L_\nu L_\sigma L_\tau)
\end{equation}
to which the associated charge, $B=\int B^0$, is taken as the baryon number of $U$ (and is identical to the degree discussed above). Both points of view show that $\partial_0 B=0$.

$B$ classifies the field configurations $U$ into topological sectors with different baryon numbers. Minimising the energy in one sector will achieve a static solution with a given baryon number. Quantising the rotational degrees of freedom gives spinning solutions which are the true nucleons and nuclei in this model.

\subsection{The $B=1$ Skyrmion}

The basic static non-vacuum solution of equation (\ref{eq:field}) is the so-called hedgehog solution \cite{adkins1983,manton2004}, which is spherically symmetric in the sense that $U(R(A)\vec x)=A U(\vec x) A^{-1}$, where $R(A)$ is the $\SO(3)$ matrix corresponding to the $\SU(2)$ matrix $A$ under the canonical $\SU(2)\rightarrow \SO(3)$ (two-to-one) homomorphism. The hedgehog solution is believed to be the Skyrmion with baryon number one, i.e. the energy minimiser in the $B=1$ sector. The hedgehog solution is of the form
\begin{equation}
U(\vec x) = \exp\left(i f(|\vec x|) \hat x \cdot \vec \tau \right)
\end{equation}
where $\hat x=\vec x / |\vec x|$ and $f(r)$ is the profile function. W.l.o.g. $f(\infty)=0$ so that $U(\infty)=1$ (any rotation in real or isospin space of this $U$ would still be spherically symmetric, although it would not technically be a hedgehog, which was chosen as a name as all the pion fields are pointing outwards).

By using the spherical symmetry, one can find the differential equation $f$ has to obey in order to be a static solution. This is
\begin{equation}\label{eq:spher}
\left(r^2 + 2 \sin^2 f\right)f''+2rf'+\sin 2f \left(f'^2-1-\frac{\sin^2f}{r^2}\right)-m^2 r^2 \sin f=0 
\end{equation}
with the boundary conditions $f(0)=\pi$ and $f(\infty)=0$.

Eq. (\ref{eq:spher}) can be solved numerically and the spinning Skyrmions have been analysed by Adkins, Nappi and Witten (massless pion case) \cite{adkins1983} and Adkins and Nappi (massive pions) \cite{adkins1984} to find the properties of nucleons and delta resonances.

Linearising the hedgehog equation (\ref{eq:spher}) yields
\begin{equation}\label{eq:skasym}
r^2 f'' + 2 r f' - 2 f - m^2 r^2 f = 0
\end{equation}
which is the modified spherical Bessel equation of first order. This means, considering the asymptotic pion fields as massive scalar fields, the hedgehog assumption forces an asymptotic dipole field (the same is true in the massless case), given by a rescaled first order modified spherical Bessel function of the second kind (which approaches $0$ when $r\rightarrow\infty$)
\begin{equation}
f(r) \sim C_1 \e^{-mr} \left(\frac{1}{r^2}+\frac{m}{r}\right) \text{ .} \label{eq:c1asym}
\end{equation}
In the massless case, it can actually be shown that for any static Skyrmion the lowest-order nonzero multipole is at least a dipole. It cannot have an asymptotic monopole \cite{manton1994}. It is not known whether this result holds in the massive pion case, however a monopole is ruled out due to symmetry for the hedgehog.

Actually, there are three massive pion fields. The hedgehog has a dipole falloff in each of these components, the dipole moments being
\begin{equation}
4\pi\left(\begin{matrix}C_1\\0\\0\end{matrix}\right) \text{, }4\pi\left(\begin{matrix}0\\C_1\\0\end{matrix}\right) \text{, } 4\pi\left(\begin{matrix}0\\0\\C_1\end{matrix}\right) \text{ ,}
\end{equation}
respectively, for the $\pi_1$, $\pi_2$ and $\pi_3$ components. They can be put together in one matrix, which is a multiple of the identity matrix: $4\pi C_1 I_3$. This is really useful in case of rotated hedgehogs: Rotated by the $\SO(3)$ rotation matrix $R$, the dipole moment matrix becomes $4\pi C_1 R$. The general asymptotic interaction energy of two rotated hedgehogs, 
\begin{align}
E_\mathrm{int} =& \frac{2 C_1^2}{9 \pi} \left( \left( \frac{m^2}{|\vec X|}+\frac{3m}{|\vec X|^2}+\frac{3}{|\vec X|^3}\right) \e^{-m|\vec X|} \left(3 R^{\color{DarkRed}{(1)}}_{ik} \hat X_k R^{\color{DarkBlue}{(2)}}_{ij} \hat X_j- R^{\color{DarkRed}{(1)}}_{ij} R^{\color{DarkBlue}{(2)}}_{ij}\right) \right.\nonumber\\
&\left. + \frac{m^2 \e^{-m|\vec X|}}{|\vec X|} R^{\color{DarkRed}{(1)}}_{ij} R^{\color{DarkBlue}{(2)}}_{ij} \right) \text{ ,}\label{eq:classint}
\end{align}
is given in \cite{jackson1985}. Here $|\vec X|$ is the distance between the two Skyrmions, $R^{\color{DarkRed}{(1)}}$ and $R^{\color{DarkBlue}{(2)}}$ are the rotation matrices (relative to the hedgehog), and $\hat X$ is the unit vector pointing in the direction of separation.

Using numerics, the constant $C_1$ is easily accessible. The method employed is the same as is used later to find asymptotic coefficients in more difficult cases, but in this case we only need to solve an ordinary differential equation (ODE). A shooting method is applied, choosing a value for $C_1$ (see eq. (\ref{eq:c1asym})) first. Then, starting from some finite point $L$ ($L$ is chosen large so that the asymptotic is valid at $L$), ODE solution methods are used to find the function $f(r)$ numerically, by solving the equation from right to left.\footnote{Using the {\tt odeint} method from Python's {\tt SciPy} package \cite{scipy}} The value at $f(0)$ is then compared with the target value $\pi$, and $C_1$ adjusted until it is attained with a given precision. Shooting from right to left circumvents the difficulty of deciding whether a solution goes to zero at infinity. This is particularly difficult for this equation, because missing zero by only a tiny bit, the solution will always escape to infinity as $r\rightarrow\infty$. \label{sec:c1det}

\begin{figure}[!bp]
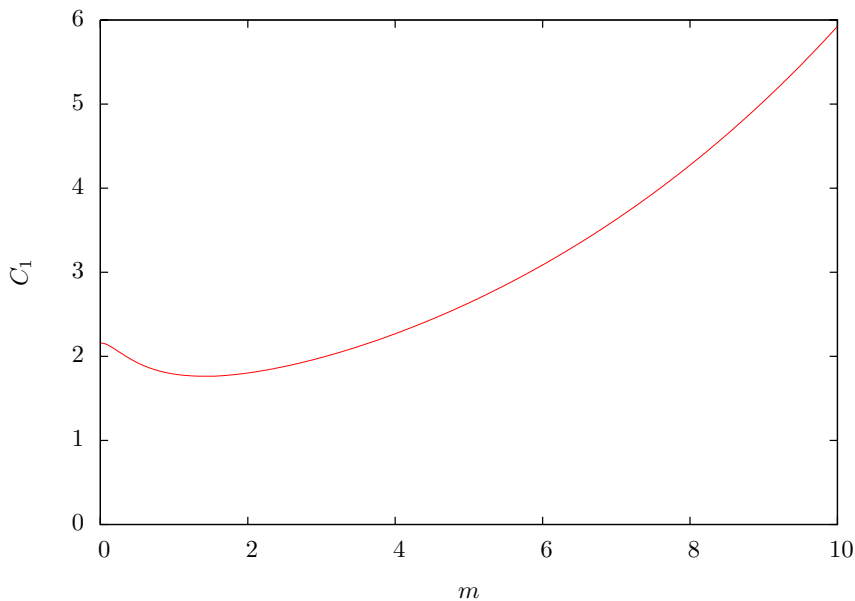

        \centering
        \resizebox{0.9\textwidth}{!}{\input massparams10.tex}
        \caption{Dependence of the hedgehog dipole parameter $C_1$ on the pion mass $m$.}
        \label{fig:massparams}
\end{figure}

In figure \ref{fig:massparams} the value of $C_1$ is plotted against the pion mass parameter $m$. It is quite surprising that $C_1$ has a minimum at positive $m$. The value $C_1=2.16$ at $m=0$ was known previously \cite{manton2004}.

\section{The interaction energy of Yukawa multipoles}\label{sec:scalint}

\begin{figure}[!bp]
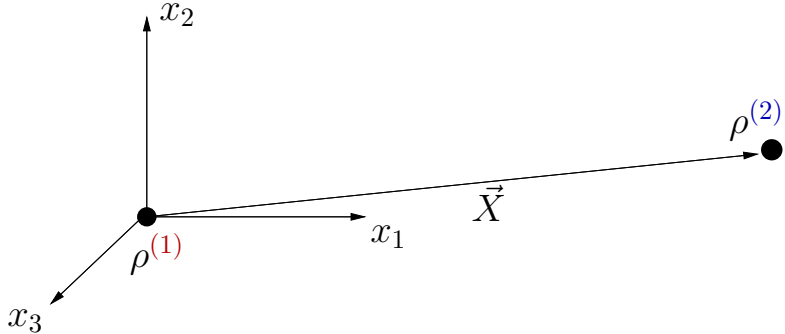

        \centering
         \input genericmultisetting.pdftex_t
        \caption{Geometry of the two-multipole computation.}
        \label{fig:genericmultisetting}
\end{figure}

In this section we derive the expression for the interaction energy of a $2^l$-pole with a $2^m$ pole in a massive scalar (Yukawa) field theory, needed in the following sections to compute Skyrmion interaction energies. Although the expression is simple, it does not seem to be well known in this generality.

A massive scalar field theory with source $\rho$ is given by the energy
\begin{equation}
E = \int \left( \frac{1}{2} (\partial_i \phi)(\partial_i \phi) + \frac{1}{2} m^2 \phi^2 + \rho \phi\right) \de^3 x \label{eq:scfield}
\end{equation}
which leads to the linear field equation
\begin{equation}
\left(\vec \nabla^2 - m^2\right) \phi = \rho \label{eq:scalfieldeq} \text{ .}
\end{equation}

The charge distribution of two Yukawa multipoles can be written as
\begin{equation}
\rho(\vec x) =(-1)^l Q^{\color{DarkRed}{(1)}}_{i_1 \ldots i_l} (\partial_{i_1} \ldots \partial_{i_l} \delta) (\vec x) + (-1)^m Q^{\color{DarkBlue}{(2)}}_{i_1 \ldots i_m} (\partial_{i_1} \ldots \partial_{i_m} \delta) (\vec x- \vec X)
\end{equation}
where the multipoles are separated by $\vec X$ (see figure \ref{fig:genericmultisetting}). As our field theory is linear, the potential is simply the sum of the two multipole potentials
\begin{equation}
\phi(\vec x) = (-1)^l Q^{\color{DarkRed}{(1)}}_{i_1 \ldots i_l} (\partial_{i_1} \ldots \partial_{i_l} \varphi) (\vec x) + (-1)^m Q^{\color{DarkBlue}{(2)}}_{i_1 \ldots i_m} (\partial_{i_1} \ldots \partial_{i_m} \varphi) (\vec x- \vec X) \text{ .}
\end{equation}
Here $\varphi(\vec x)=-\e^{-m|\vec x|}/(4\pi |\vec x|)$ is the Green's function of the massive scalar field theory.

The interaction energy is the difference of the total energy and the energies of the individual multipole fields, i.e. $E_\text{int} = E(\phi,\rho) - E(\phi^{\color{DarkRed}{(1)}},\rho^{\color{DarkRed}{(1)}}) - E(\phi^{\color{DarkBlue}{(2)}},\rho^{\color{DarkBlue}{(2)}})$. Generally, the energy of point charges and all their derivatives is infinite, however, the difference $E_\text{int}$ is finite, as can be seen from the following equation which is easily derived from the original expression (eq. (\ref{eq:scfield})):
\begin{align}
E(\phi,\rho) =& E(\phi^{\color{DarkRed}{(1)}},\rho^{\color{DarkRed}{(1)}}) + E(\phi^{\color{DarkBlue}{(2)}},\rho^{\color{DarkBlue}{(2)}}) \nonumber\\&+ \int \de^3 x \left(  (\partial_i \phi^{\color{DarkRed}{(1)}})(\partial_i \phi^{\color{DarkBlue}{(2)}}) + m^2 \phi^{\color{DarkRed}{(1)}} \phi^{\color{DarkBlue}{(2)}} + \rho^{\color{DarkRed}{(1)}} \phi^{\color{DarkBlue}{(2)}} + \rho^{\color{DarkBlue}{(2)}} \phi^{\color{DarkRed}{(1)}}\right) \text{ .}
\end{align}
The last integral is therefore the interaction energy. Using the field equations, this can be reduced to
\begin{equation}
E_\text{int,Yuk} = \int \de^3 x \rho^{\color{DarkRed}{(1)}} \phi^{\color{DarkBlue}{(2)}} \text{ .}
\end{equation}
We can as well write $E_\text{int,Yuk} = \int \de^3 x \rho^{\color{DarkBlue}{(2)}} \phi^{\color{DarkRed}{(1)}}$ by exchanging the roles of the two fields. Using our expressions for $\rho^{\color{DarkBlue}{(2)}}$ and $\phi^{\color{DarkRed}{(1)}}$, we get
\begin{align}
E_\text{int,Yuk} = (-1)^l Q^{\color{DarkBlue}{(2)}}_{i_1 \ldots i_m} Q^{\color{DarkRed}{(1)}}_{j_1 \ldots j_l}  (\partial_{i_1} \ldots \partial_{i_m} \partial_{j_1} \ldots \partial_{j_l} \varphi) (\vec X) \text{ .} \label{eq:aeint}
\end{align}
This is a surprisingly simple and symmetric result. In order to find the interaction energy between any two multipoles, we just need to calculate the appropriate derivative of the potential (Green's) function. Note that compared to the interaction energy of the Coulomb potential
\begin{align}
E_\text{int,Coulomb} = (-1)^{l+1} Q^{\color{DarkBlue}{(2)}}_{i_1 \ldots i_m} Q^{\color{DarkRed}{(1)}}_{j_1 \ldots j_l}  (\partial_{i_1} \ldots \partial_{i_m} \partial_{j_1} \ldots \partial_{j_l} ) \left(-\frac{1}{4\pi|\vec X|}\right)
\end{align}
which is derived in \cite[eq.~(8)]{jansen1958}, equation (\ref{eq:aeint}) here differs by a sign and the different choice of potential. The opposite sign is due to the Skyrme theory being a theory of scalar interactions, whereas electromagnetic interactions are vector interactions.

\subsection{Yukawa theory for asymptotic interactions}

In this section, we are going to show that to find asymptotic interactions of Skyrmions, it is sufficient to understand Yukawa multipole theory and know the falloff of the Skyrmions involved to get the complete picture. 

We compute the interaction energy using the product ansatz: given two Skyrmions defined by their Skyrme fields $U^{\color{DarkRed}{(1)}}$ and $U^{\color{DarkBlue}{(2)}}$, we approximate their combined fields by $U=U^{\color{DarkRed}{(1)}}U^{\color{DarkBlue}{(2)}}$. The interaction energy is then defined by the equation
\begin{equation}\label{eq:eintdef}
E(U) = E(U^{\color{DarkRed}{(1)}})+E(U^{\color{DarkBlue}{(2)}})+E_\mathrm{int} \text{ .}
\end{equation}
It is then useful to write the $\SU(2)$ fields in terms of their logarithms, i.e.
\begin{align}
U^{\color{DarkRed}{(1)}}(\vec x) = &\exp\left(i F^{\color{DarkRed}{(1)}}_j (\vec x) \tau_j\right) \label{eq:fidef1} \\ 
U^{\color{DarkBlue}{(2)}}(\vec x) = &\exp\left(i F^{\color{DarkBlue}{(2)}}_j(\vec x) \tau_j\right) \label{eq:fidef2}  \text{ .}
\end{align}

For convenience, we will assume our Skyrmions are centered at $(0,0,0)$ for $U^{\color{DarkRed}{(1)}}$ and $\vec X=(X,0,0)$ for $U^{\color{DarkBlue}{(2)}}$, where $X$ is large.

Let us write down the expression for the total energy $E(U)$: We will immediately split up the energy into two integrals over the domains $\Omega^{\color{DarkRed}{(1)}} = \{\vec x \in \IR^3 | x_1 < X/2\}$ and $\Omega^{\color{DarkBlue}{(2)}} = \{\vec x \in \IR^3 | x_1 \geq X/2\}$ and linearise $U^{(n)}=1+iF^{(n)}_j(\vec x) \tau_j$ where appropriate. Using integration by parts, this yields
\begin{align}
&E(U) \approx E(U^{\color{DarkRed}{(1)}})+E(U^{\color{DarkBlue}{(2)}})\\
&+\frac{1}{12 \pi^2}\int_{\Omega^{\color{DarkRed}{(1)}}} \de^3 x \Tr \left( i F^{\color{DarkBlue}{(2)}}_k \tau_k \left( \partial_i\left(L^{\color{DarkRed}{(1)}}_i + \frac{1}{4}[L^{\color{DarkRed}{(1)}}_j,[L^{\color{DarkRed}{(1)}}_j,L^{\color{DarkRed}{(1)}}_i]]\right) + m^2 U^{\color{DarkRed}{(1)}}\right)\right)\nonumber\\
&+\frac{1}{12 \pi^2}\int_{\Omega^{\color{DarkBlue}{(2)}}} \de^3 x \Tr \left( i F^{\color{DarkRed}{(1)}}_k \tau_k \left( \partial_i\left(L^{\color{DarkBlue}{(2)}}_i + \frac{1}{4}[L^{\color{DarkBlue}{(2)}}_j,[L^{\color{DarkBlue}{(2)}}_j,L^{\color{DarkBlue}{(2)}}_i]]\right)  + m^2 U^{\color{DarkBlue}{(2)}}\right)\right)\nonumber\\
&-\frac{1}{12 \pi^2}\int_{\{X/2\} \times \IR^2} \de x_2 \de x_3 \Tr \left( i F^{\color{DarkBlue}{(2)}}_k \tau_k \left(L^{\color{DarkRed}{(1)}}_1 + \frac{1}{4}[L^{\color{DarkRed}{(1)}}_j,[L^{\color{DarkRed}{(1)}}_j,L^{\color{DarkRed}{(1)}}_1]]\right) \right) \nonumber\\
&+\frac{1}{12 \pi^2}\int_{\{X/2\} \times \IR^2} \de x_2 \de x_3 \Tr \left( i F^{\color{DarkRed}{(1)}}_k \tau_k \left(L^{\color{DarkBlue}{(2)}}_1 + \frac{1}{4}[L^{\color{DarkBlue}{(2)}}_j,[L^{\color{DarkBlue}{(2)}}_j,L^{\color{DarkBlue}{(2)}}_1]]\right) \right)\nonumber
\text{ .}
\end{align}
We can eliminate all the volume integrals using the field equations (\ref{eq:field}), and what remains is 
\begin{equation}
E_\mathrm{int} = \frac{1}{6 \pi^2}\int_{\{X/2\} \times \IR^2} \de x_2 \de x_3 \left(F^{\color{DarkBlue}{(2)}}_i \partial_1 F^{\color{DarkRed}{(1)}}_i - F^{\color{DarkRed}{(1)}}_i \partial_1 F^{\color{DarkBlue}{(2)}}_i\right) \label{eq:planeint}
\end{equation}
to linear order (this equation was first mentioned in the context of the Skyrme model in \cite{mantonschroers2004}). Doing the same computation in the massive scalar field theory defined by equation (\ref{eq:scalfieldeq}), i.e. computing the interaction energy of two charge distributions centered at $(0,0,0)$ and $(X,0,0)$, again using the field equations to reduce it to a planar integral, yields
\begin{equation}
E_\mathrm{int,Yuk} = \int_{\{X/2\} \times \IR^2} \de x_2 \de x_3 \left(\phi^{\color{DarkBlue}{(2)}} \partial_1 \phi^{\color{DarkRed}{(1)}} - \phi^{\color{DarkRed}{(1)}} \partial_1 \phi^{\color{DarkBlue}{(2)}}\right) \label{eq:yukplane}
\end{equation}
with $\phi^{(n)}$ generated by the charge distribution $\rho^{(n)}$.

This means to linear order (which is correct asymptotically, i.e. for $X\rightarrow\infty$ if the fields fall off quickly) the interactions are described by the massive scalar field theory. Thus, given a $2^l$- and a $2^m$-pole in the Skyrme theory, their interaction for general separation $\vec X$ is
\begin{align}
E_\text{int} = \frac{1}{6\pi^2} (-1)^l Q^{i,\color{DarkBlue}{(2)}}_{i_1 \ldots i_m} Q^{i,\color{DarkRed}{(1)}}_{j_1 \ldots j_l}  (\partial_{i_1} \ldots \partial_{i_m} \partial_{j_1} \ldots \partial_{j_l} \varphi) (\vec X) \label{eq:skyrmemultiint}
\end{align}
where the different prefactor of the integrals in equations (\ref{eq:planeint}) and (\ref{eq:yukplane}) has been taken into account, as well as that there are three scalar fields $F_i$ in the Skyrme theory.

\section{The interactions of two $B=4$ Skyrmions}

\subsection{The $B=4$ solution}

\begin{figure}[ht]
        \centering
	\begin{minipage}{0.25\linewidth}\hspace{1cm}
	\resizebox{0.6\textwidth}{!}{\input coord_1.pdftex_t}
	\end{minipage}
	\begin{minipage}{0.7\linewidth}
        \includegraphics[width=7cm]{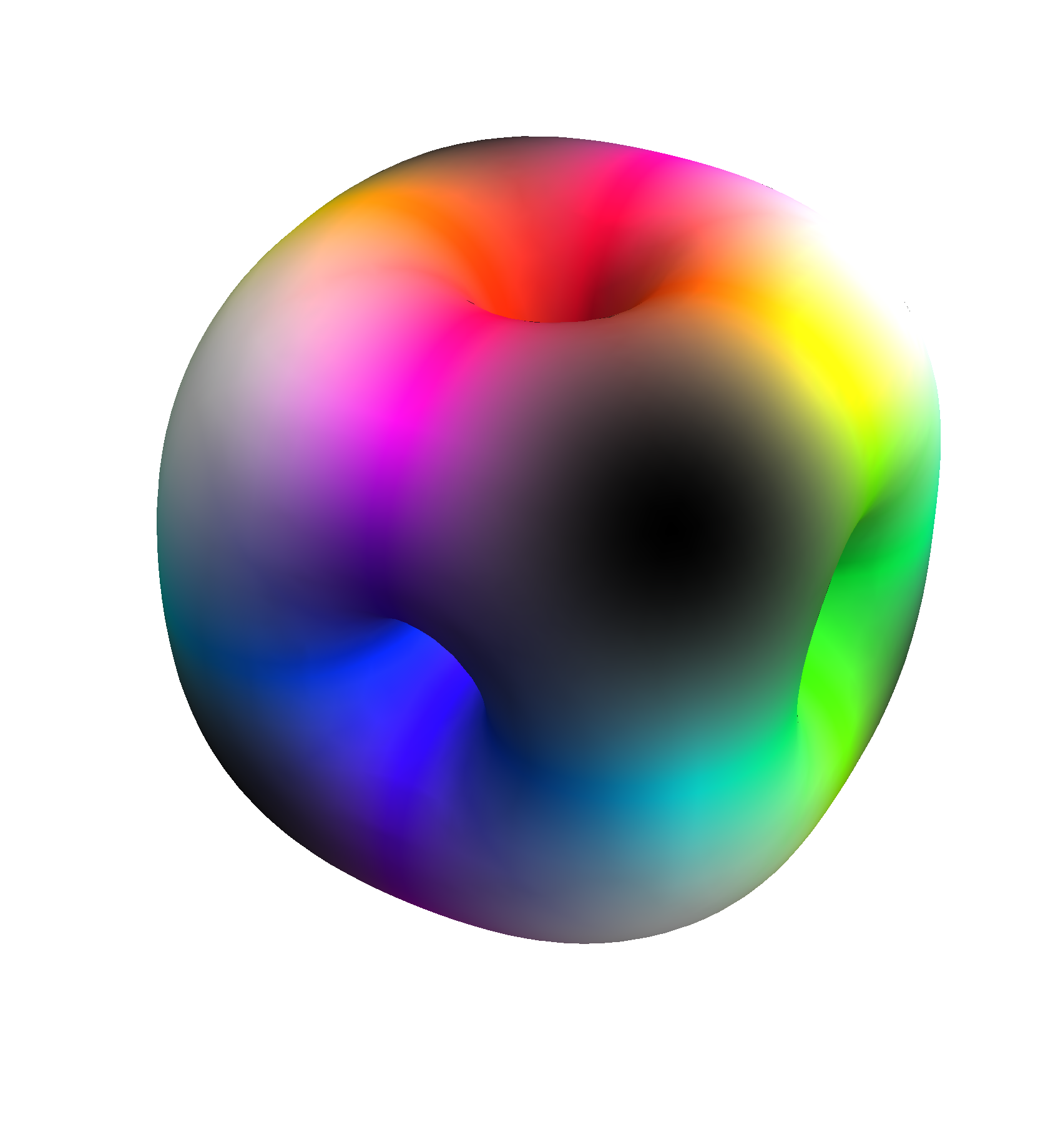}
	\end{minipage}
	\caption{The $B^0=0.1$ isobaryon surface of the $B=4$ solution in its standard orientation. The colours indicate the direction of the pion fields (see figure \ref{fig:coloursphere}). Opposite faces of the $B=4$ solution have the same colour.}
	\label{fig:alpha}
\end{figure}

The Skyrmion solution with baryon number 4 (figure \ref{fig:alpha}), first given in \cite{braaten1990}, plays an outstanding role in the Skyrme model. When quantised, it serves as the model for nuclei with four nucleons, notably the ${}^4 \mathrm{He}$ nucleus or $\alpha$-particle, but also ${}^4 \mathrm{H}$ and ${}^4 \mathrm{Li}$, which are however only seen as resonances and not as stable nuclei \cite{manko2007}.

Interestingly, heavier nuclei seem to be composed mostly of interacting clusters of $\alpha$-particles, a fact known to experimentalists and understood to a certain extent in the Skyrme model \cite{battye2009}. Nuclei with a baryon number divisible by four tend to be particularly well described by an arrangement of $\alpha$-particles without any remaining single nucleons, and this is where the Skyrme model has been the most successful to date. This is why we are interested in knowing how the $B=4$ solution interacts with copies of itself and other solutions of the Skyrme model.

The $B=4$ solution looks like a cube and has cubic symmetry. We will often be referring to the standard orientation of this solution in this paper. This is the orientation arising when one constructs the solution starting from the rational map
\begin{equation}
R(z) = \frac{z^4+2\sqrt{3}iz^2+1}{z^4-2\sqrt{3}iz^2+1} \label{eq:ratmap}
\end{equation}
using the rational map formalism described in \cite{houghton1997}. In this formalism, the rational map encodes the angular dependence of the Skyrme field, and an additional function $f(r)$ gives the radial dependence.

\subsection{The symmetry group}

Let us describe the cubic symmetry of the $B=4$ solution. The symmetry acts as the $O_h$ subgroup of $\Oh(3)$ on the spatial coordinates \cite{manko2007}, which is also known as the $T_{1u}$ representation of $O_h$.\footnote{We are using the notation from \cite{cotton1993} for irreducible representations of $O_h$. One-dimensional representations are labeled $A_1$, $A_2$, two-dimensional $E$ and three-dimensional $T_1$, $T_2$. Additionally, the index $g$ ({\it gerade}=even) marks an even representation (where the inversion of $O_h$ is mapped to the identity) and $u$ ({\it ungerade}=odd) an odd representation (where it is mapped to the negative identity).} It does not act trivially on the Skyrme field. The action is easiest to describe as a group representation on the pion $F_1$, $F_2$ and $F_3$ fields. In the standard orientation the $O_h$ symmetry can then be deduced from the rational map (eq. (\ref{eq:ratmap})) as acting by the $E_g$ representation on the $(F_1,F_2)$ field components and by $A_{2u}$ on $F_3$.

\begin{figure}[hbp]
	\begin{minipage}{0.5\linewidth}
	\begin{minipage}{0.33\linewidth}\hspace{0.7cm}
	\resizebox{0.6\textwidth}{!}{\input coord_1pi.pdftex_t}
	\end{minipage}
	\begin{minipage}{0.62\linewidth}
	\includegraphics[width=3.5cm]{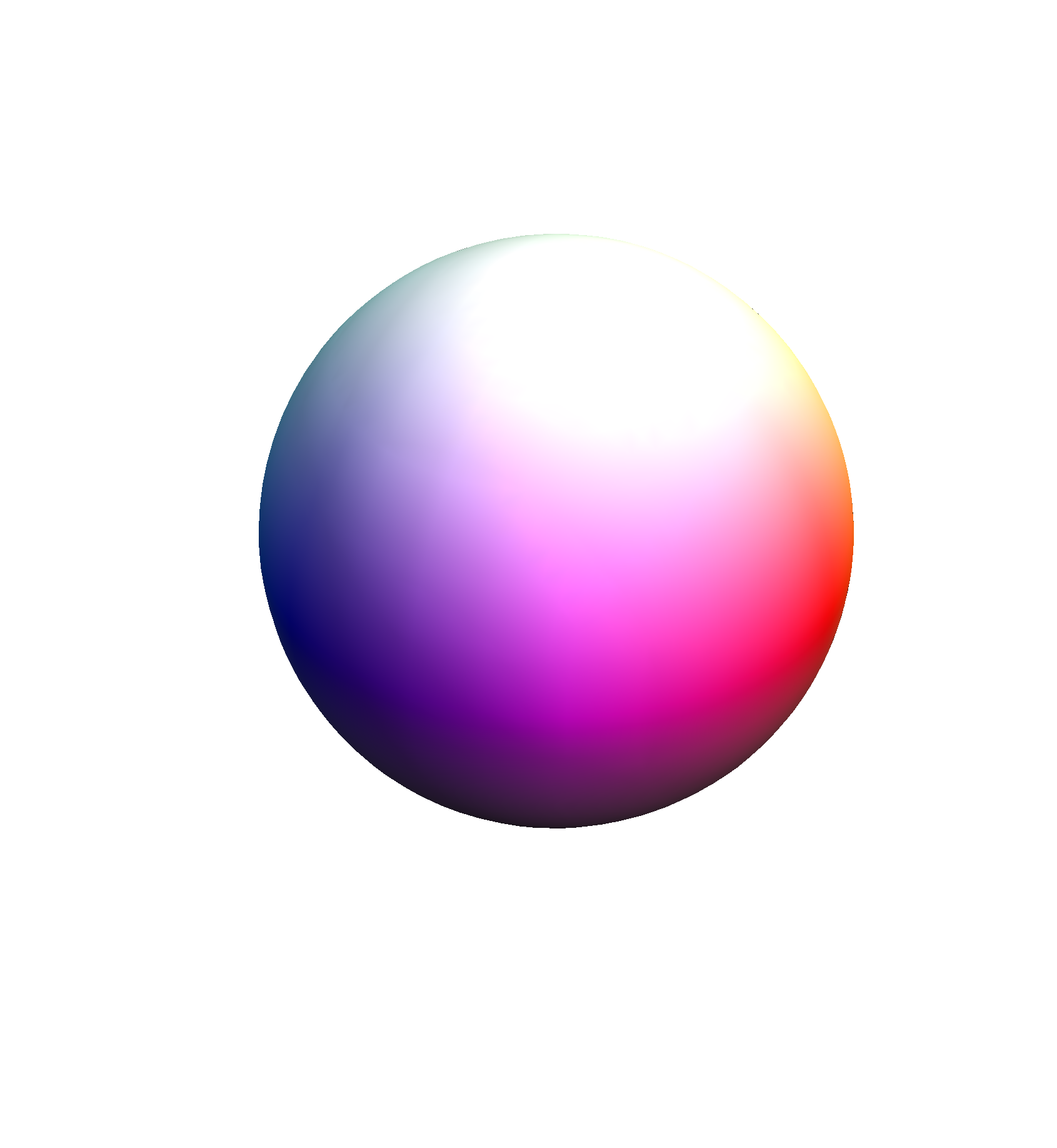}	
	\end{minipage}
	\end{minipage}
	\vline \hspace{0.5cm}
	\begin{minipage}{0.5\linewidth}
	\begin{minipage}{0.33\linewidth}
	\resizebox{0.9\textwidth}{!}{\input coord_2.pdftex_t}
	\end{minipage}
	\begin{minipage}{0.62\linewidth}
	\includegraphics[width=3.5cm]{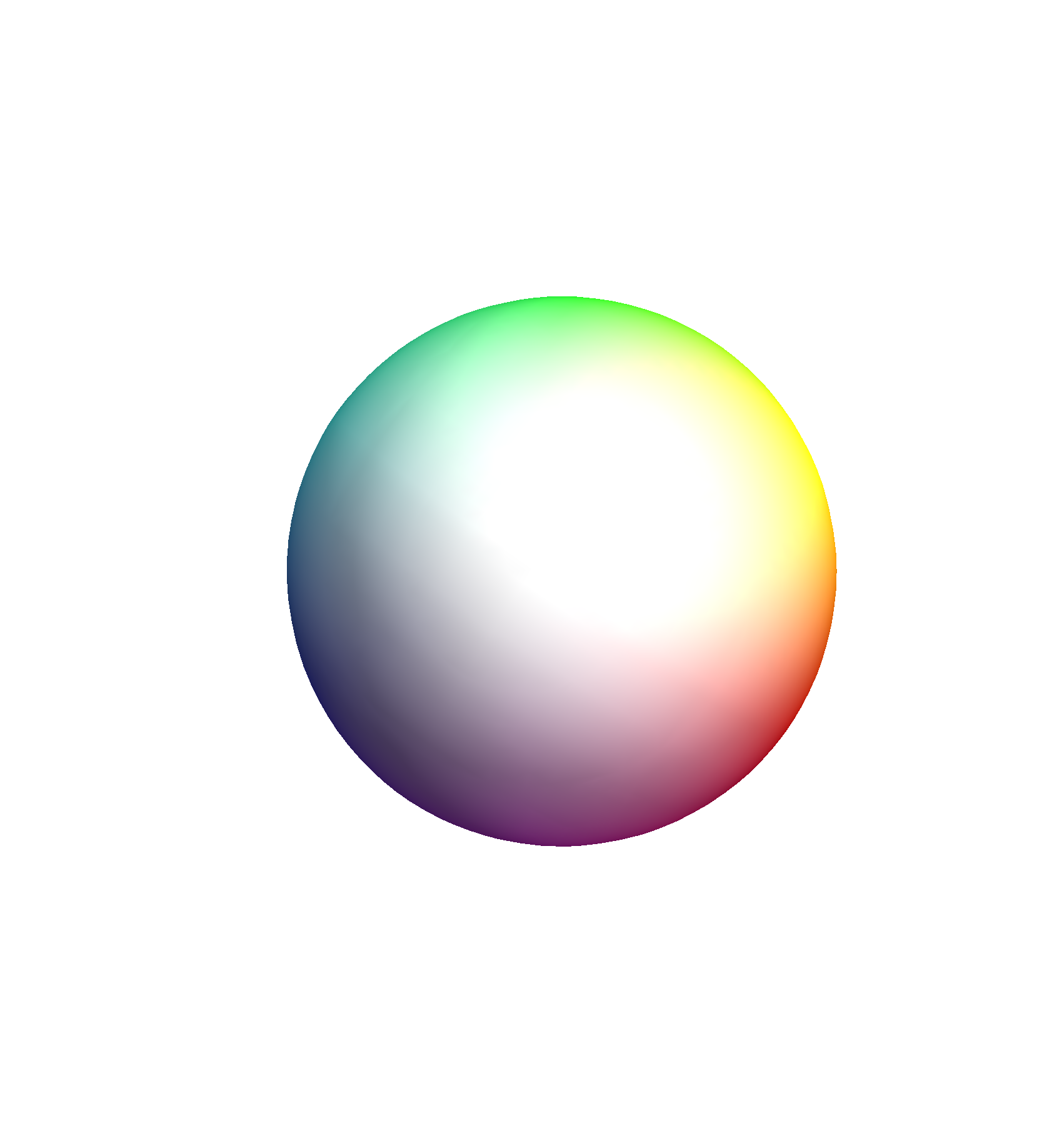}
	\end{minipage}
	\end{minipage}
	\caption{The colour sphere, used to colour the pion fields, in two different orientations (the second, top view shows the colour circle in the $(\pi_2,\pi_2)$-plane). An isobaryon surface of the $B=1$ Skyrmion with coloured pion fields will have the same appearance as the colour sphere.}
	\label{fig:coloursphere}
\end{figure}

There is a colour scheme corresponding particularly well to this symmetry, using Runge's colour sphere to indicate the direction of the pion fields $\vec \pi / |\vec \pi|$. The equator, the unit circle of the $(\pi_1,\pi_2)$ plane, is coloured using the hue colour attribute: This construction is also known as the colour circle. The $\pi_3$ direction is indicated by lightness, making the north pole ($\pi_3=+1$) white and the south pole ($\pi_3=-1$) black -- see figure \ref{fig:coloursphere}. In this colour scheme, the corners of the $B=4$ cube in standard orientation are black and white, whereas the faces are associated with the three fundamental colours green, blue and red.

\subsection{$B=4$ quadrupole and octupole tensors}

We want to apply the Yukawa multipole formalism to find the interactions of two $B=4$ Skyrmions, so we need to identify the dominant pion multipoles of each of these. To this end, we are going to use the high symmetry of the Skyrmion to reduce the task as much as possible. 

We have three multipole expansions, one for each $F_i$ field, whose multipole tensors we denote by $Q^i_{i_1\ldots i_m}$. We know how a multipole transforms under $\Oh(3)$ as a tensor by applying the $\Oh(3)$ matrix to each $i_j$ index. $O_h$, as a subgroup of $\Oh(3)$, will act accordingly. However, there is another way to do the same transformation: doing the corresponding isospace transformation. Isospin transformations act on the $i$ index. This gives us strong compatibility requirements, which we are going to explore next.

As noted earlier, no static solution of the Skyrme field equation without pion mass can have a monopole. This is not true in the massive pion case. A monopole, being a scalar, will transform according to the $A_{1g}$ representation of $O_h$. However $(F_1,F_2)$ transforms according to $E_g$, and $F_3$ according to $A_{2u}$, so there is no monopole, even in the massive pion case.

The next multipole to check is the dipole, which transforms according to the $\mathbf{1}_u$ representation\footnote{We use $\mbox{\boldmath$\ell$}_{g/u}$ for representations of $\Oh(3)$, where $\ell$ is the angular momentum associated to the representation and the $g$/$u$ index denotes even and odd, the same as used in $O_h$ representations.} of $\Oh(3)$, which corresponds to the $T_{1u}$ vector representation of $O_h$. But under our isospace action, neither the $(F_1,F_2)$-fields nor $F_3$ include any $T_{1u}$ part, so the dipole can only be zero.

The quadrupole, in our definition, transforms according to the $\mathbf{2}_g \oplus \mathbf{0}_g$ representation\footnote{It should be noted that this includes a trace part, whereas the common definition is for the quadrupole tensor in massless theories (e.g. electrodynamics) to be traceless. This however cannot be guaranteed in a massive field theory, and therefore we cannot a priori exclude trace parts in this and higher order multipole tensors.} of $\Oh(3)$, which corresponds to $E_g \oplus T_{2g} \oplus A_{1g}$ of $O_h$. But we know $O_h$ acts as $A_{2u}$ on $F_3$, so this field does not have a quadrupole moment. However, the $(F_1,F_2)$ doublet transforms as $E_g$, and $E_g$ is in the quadrupole representation -- so it is possible to have a quadrupole in these fields.

An explicit computation gives the following form for the quadrupole tensors:
\begin{eqnarray}
Q^1 &=& 4 \pi C_2 \ \ \ \left(\begin{matrix}
-1 & 0 & 0 \\
0 & -1 & 0 \\
0 & 0 & \ 2\ 
\end{matrix}\right) \nonumber \\
Q^2 &=& 4 \pi C_2 \left(\begin{matrix}
\ \sqrt{3}\  & 0 & 0 \\
0 & -\sqrt{3}  & 0 \\
0 & 0 & \ \,0\ \,
\end{matrix}\right) \label{eq:quadrutens} \text{ .}
\end{eqnarray}
This basis of traceless quadrupoles obviously depends on the orientation and is given in the standard orientation we are using. The single constant $C_2$ is left to be determined by numerics.

The octupole transforms according to $\mathbf{3}_u \oplus 3 \cdot \mathbf{1}_u$ in $\Oh(3)$, which becomes $A_{2u} \oplus 4 T_{1u} \oplus T_{2u}$ in $O_h$. This representation has no $E_g$ part, so the $F_1$ and $F_2$ octupoles are forced to zero. There is now, however, an $A_{2u}$ part, so we can have an octupole in $F_3$.

This octupole tensor can be found to be
\begin{equation}
Q^3_{ijk} = \frac{4 \pi C_3}{3!} |\varepsilon_{ijk}| \label{eq:octens}
\end{equation}
where we write $ |\varepsilon_{ijk}|$ for a tensor that is nonzero in the same components as the totally antisymmetric tensor, only that it takes the value $1$ in all of these.

\subsection{The three-colour decomposition}

Looking at the $B=4$ solution in figure \ref{fig:alpha} we see a very high degree of symmetry, reflected also in the large symmetry group $O_h$. Despite this, at first sight the quadrupole tensors we found in equations (\ref{eq:quadrutens}) do not look very symmetrical in the axes.

The reason for the apparent asymmetry of these tensors becomes clear when we look at the colour circle (figure \ref{fig:coloursphere}): In it, the three colours appearing on the faces of the $B=4$ cube occur separated by angles of $2\pi/3=120^\circ$; however we are pressing them into the form of two orthogonal axes along $F_1$ and $F_2$. Would it be possible to rewrite the interaction in the form of three quadrupoles? The answer is yes, and the result is very neat and useful for us to get closer to a quadrupole interaction that is easy to work with.

From equation (\ref{eq:skyrmemultiint}), we infer that the interaction of the two pairs of quadrupoles is given by
\begin{equation}
E_\text{int}^{l=2} = Q^{1,\color{DarkRed}{(1)}}_{ij} M_{ijkl} Q^{1,\color{DarkBlue}{(2)}}_{kl} + Q^{2,\color{DarkRed}{(1)}}_{ij} M_{ijkl} Q^{2,\color{DarkBlue}{(2)}}_{kl} \label{eq:twoquadruint} 
\end{equation}
where we introduce the quadrupole interaction tensor $M_{ijkl} = \frac{1}{6\pi^2} (\partial_{i} \partial_{j} \partial_{k} \partial_{l} \varphi) (\vec X)$.

Now we consider splitting $Q^1$ and $Q^2$ into three quadrupoles, one for each colour. Let us give the ``green'', ``blue'' and ``red'' quadrupole tensors numbers 1, 2 and 3 respectively and label these quadrupoles $P^c$. Due to the $120^\circ$ angles between colours, our decomposition should work like this:
\begin{eqnarray}
\frac{1}{K} P^{3} &=& Q^{1} \nonumber\\
\frac{1}{K} P^{1} &=& -\frac{1}{2} Q^{1}+\frac{\sqrt{3}}{2}Q^{2} \\
\frac{1}{K} P^{2} &=& -\frac{1}{2} Q^{1}-\frac{\sqrt{3}}{2}Q^{2} \nonumber
\end{eqnarray}
where $K$ is a constant. Note that $P^1+P^2+P^3=0$. We find that using $K^2 = \frac23$ the interaction energy in eq. (\ref{eq:twoquadruint}) becomes
\begin{equation}
E_\text{int}^{l=2} =P^{1,\color{DarkRed}{(1)}}_{ij} M_{ijkl} P^{1,\color{DarkBlue}{(2)}}_{kl} + P^{2,\color{DarkRed}{(1)}}_{ij} M_{ijkl} P^{2,\color{DarkBlue}{(2)}}_{kl} + P^{3,\color{DarkRed}{(1)}}_{ij} M_{ijkl} P^{3,\color{DarkBlue}{(2)}}_{kl}\label{eq:threetwoquad} \text{ .}
\end{equation}
This leads to an intuitive picture of one interaction term for each colour (green, blue and red).

Assume for a moment that the Skyrmion is unrotated, giving $Q^{1}$ and $Q^{2}$ the form in eq. (\ref{eq:quadrutens}). The three colour quadrupole tensors are
\begin{eqnarray}
P^{1} &=& \sqrt\frac23 4 \pi C_2 \left(\begin{matrix}
2 & 0 & 0 \\
0 & -1 & 0 \\
0 & 0 & -1
\end{matrix}\right) \nonumber\\
P^{2} &=& \sqrt\frac23 4 \pi C_2 \left(\begin{matrix}
-1 & 0 & 0 \\
0 & 2 & 0 \\
0 & 0 & -1
\end{matrix}\right) \\
P^{3} &=& \sqrt\frac23 4 \pi C_2 \left(\begin{matrix}
-1 & 0 & 0 \\
0 & -1 & 0 \\
0 & 0 & 2
\end{matrix}\right) \nonumber\text{ .}
\end{eqnarray}

\begin{figure}[tbp]
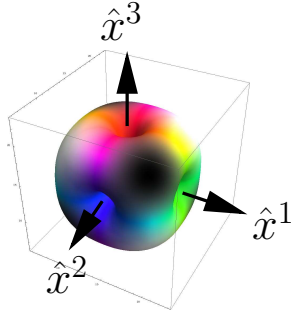

        \centering
        \input alphaqvectors.pdftex_t
        \caption{The quadrupoles $P^1$, $P^2$ and $P^3$ can be visualised as associated with vectors orthogonal to the faces of the corresponding colour.}
        \label{fig:alphaqvectors}
\end{figure}

The form of these tensors is so simple that it lends itself to another interpretation: Note that the faces coloured green point in the $x_1$ and $-x_1$ direction. Blue has the same association with the $x_2$-axis and red with the $x_3$-axis. Now, let us, for each colour, imagine a unit colour vector $\hat x^c$ pointing in the direction of one face of that colour (see figure \ref{fig:alphaqvectors}). It does not matter which of the two opposite faces we choose, as the vector will only occur in squared form. Then the corresponding colour's quadrupole is given by $P^{c}_{ij}= \sqrt\frac23 4 \pi C_2 (3 \hat x^{c}_i \hat x^{c}_j - \delta_{ij})$. In the standard orientation, $\hat x^{c}_i=\delta_{ic}$, so we can write
\begin{equation}
P^{c}_{ij}= \sqrt\frac23 4 \pi C_2( 3 \delta_{i\underline c} \delta_{j\underline c} - \delta_{ij})
\end{equation}
using the convention that underlined indices are not summed over.

Now, making a spatial rotation $R^{(n)}\in\SO(3)$ transforms $P^{c,(n)}_{ij}$ to 
\begin{equation}
P^{c,(n)}_{ij} =  \sqrt\frac23 4 \pi C_2R^{(n)}_{ik} R^{(n)}_{jl} \left( 3 \delta_{k\underline c} \delta_{l\underline c} - \delta_{kl} \right) = \sqrt\frac23 4 \pi C_2 (3 R^{(n)}_{i\underline c} R^{(n)}_{j\underline c} - \delta_{ij})
\end{equation}
where we have reintroduced the $n=1,2$ label for the Skyrmions at sites $\color{DarkRed}{(1)}$ and $\color{DarkBlue}{(2)}$.

\subsection{Interaction of unrotated $B=4$ Skyrmions}\label{sec:eint}

In order to show how to use our three-colour decomposition to get concrete formulae for the interaction, we consider the case where $R^{\color{DarkRed}{(1)}}=R^{\color{DarkBlue}{(2)}}=I$. Also we are only considering the standard isospace orientation because in most, if not all of the known solutions composed of several $B=4$ blocks, they are in the same isospace orientation. This could be because changing isospace orientation would mix quadrupole and octupole interactions, likely leading to a weaker total interaction energy.

In order to express the interaction energy easily, we need the $n$-th order modified spherical Bessel functions of the second kind, commonly denoted by $k_n(x)$, which satisfy the differential equation
\begin{equation}
x^2 \frac{\de^2 k_n(x)}{\de x^2} + 2x \frac{\de k_n(x)}{\de x} - (x^2+n(n+1)) k_n(x)=0 \text{ .}
\end{equation}
The function $k_n$ decays to zero at infinity and has a pole at $x=0$; it can be expressed in terms of elementary functions, using $k_0(x)=\e^{-x}/x$ and
\begin{equation}
k_{n+1}(x)=-x^n \frac{\de}{\de x} \left(\frac{k_n(x)}{x^n}\right) \text{ .}
\end{equation}
For convenience we use the rescaled functions
\begin{equation}
q_n(r)=\frac{(-m)^{n+1}}{4\pi} k_n(mr) \text{ .}
\end{equation}
Then, the $n$-th derivative of the Green's function $\varphi(\vec x)=-\e^{-m|\vec x|}/(4\pi |\vec x|)$ can be written as\footnote{Note that, if $m=0$, this can be simplified using $q_n(r)=(-1)^{n+1} (2n-1)!! r^{-(n+1)}/(4\pi)$.}
\begin{align}
(\partial_{i_1} \ldots \partial_{i_n}\varphi)(\vec x) = & q_n(|\vec x|) \hat x_{i_1} \hat x_{i_2} \ldots \hat x_{i_n} + \frac{q_{n-1}(|\vec x|)}{|\vec x|} [\delta_{i_1i_2} \hat x_{i_3} \hat x_{i_4} \ldots \hat x_{i_n} +\cdots] \nonumber\\ &+ \frac{q_{n-2}(|\vec x|)}{|\vec x|^2} [\delta_{i_1i_2} \delta_{i_3i_4} \hat x_{i_5} \hat x_{i_6} \ldots \hat x_{i_n} +\cdots] + \ldots 
\end{align}
where successive terms contain one more $\delta$-tensor. The notation $[T_{i_1i_2\ldots i_n}+\cdots]$ denotes the sum of all the non-equivalent terms obtained by permutation of the indices, e.g. $[\delta_{i_1i_2} \delta_{i_3i_4} + \cdots] = \delta_{i_1i_2} \delta_{i_3i_4} + \delta_{i_1i_3} \delta_{i_2i_4} + \delta_{i_1i_4} \delta_{i_2i_3}$. The quadrupole interaction tensor can then be expressed as
\begin{equation}
M_{ijkl}=\frac{1}{6 \pi^2} \left(q_4(|\vec X|)\hat X_i \hat X_j \hat X_k \hat X_l + \frac{q_3(|\vec X|)}{|\vec X|}[ \delta_{ij} \hat X_k \hat X_l +\cdots] + \frac{q_2(|\vec X|)}{|\vec X|^2} [\delta_{ij} \delta_{kl} +\cdots] \right) \text{ .}
\end{equation}
Contracting the indices of $M_{ijkl}$ with the quadrupole tensors $P^c$ as in eq. (\ref{eq:threetwoquad}), we find that the quadrupole interaction energy of two unrotated $B=4$ Skyrmions separated by $\vec X$ reduces to
\begin{equation}
E_\text{int}^{l=2}(\vec X) = \frac{16 C_2^2}{9}\left(q_4(|\vec X|)\left(9 \sum_{c=1}^3 \hat X_c^4-3\right)+24\frac{q_3(|\vec X|)}{|\vec X|}+36\frac{q_2(|\vec X|)}{|\vec X|^2}\right) \text{ .} \label{eq:l2int}
\end{equation}

Now we turn to the octupole interaction. Using the octupole interaction tensor
\begin{align}
M_{ijklmn}=-\frac{1}{6\pi^2} \left( q_6(|\vec X|)\hat X_i \hat X_j \hat X_k \hat X_l \hat X_m \hat X_n + \frac{q_5(|\vec X|)}{|\vec X|} [\delta_{ij} \hat X_k \hat X_l \hat X_m \hat X_n +\cdots] \right. \nonumber\\
\left. +\frac{q_4(|\vec X|)}{|\vec X|^2} [\delta_{ij} \delta_{kl} \hat X_m \hat X_n +\cdots] + \frac{q_3(|\vec X|)}{|\vec X|^3} [\delta_{ij} \delta_{kl} \delta_{mn} +\cdots]
 \right)
\end{align}
we obtain the octupole interaction energy
\begin{equation}
E_\text{int}^{l=3}(\vec X) = -\frac{8 C_2^2}{3}\left( q_6(|\vec X|) \hat X_1^2 \hat X_2^2 \hat X_3^2 + \frac{q_5(|\vec X|)}{2|\vec X|}\left(1-\sum_{c=1}^3 \hat X_c^4\right) + \frac{q_4(|\vec X|)}{|\vec X|^2}+\frac{q_3(|\vec X|)}{|\vec X|^3} \right) \text{.} \label{eq:l3int}
\end{equation}

The complete interaction energy of the Skyrmions up to octupole order is then $E_\text{int}^{l\leq3}(\vec X)=E_\text{int}^{l=2}(\vec X)+E_\text{int}^{l=3}(\vec X)$. Note that the constants $C_2$ and $C_3$ depend on $m$. In the $m=0$ (massless pion) case, it is possible to simplify the interaction energy to
\begin{align}
E_\text{int}^{l\leq3} \Bigg|_{m=0} =& \frac{12 C_2^2}{\pi} \frac{1}{|\vec X|^5} \left(-35\sum_{c=1}^{3} \hat X_c^4 + 21 \right)\nonumber\\
& + \frac{15 C_3^2}{\pi} \frac{1}{|\vec X|^7} \left( 462 \hat X_1^2 \hat X_2^2 \hat X_3^2  + 21 \sum_{c=1}^{3} \hat X_c^4 - 17 \right) \text{ .}
\end{align}

\subsection{Interaction of $B=4$ Skyrmions with $90^\circ$ relative rotation}\label{sec:eintr}

Another interesting case is where one of the $B=4$ Skyrmions is rotated by $90^\circ$ relative to the other one around one of the cube axes. If the separation is along the same axis, this will produce maximum attraction for the octupoles, however slightly less attraction for the quadrupoles than in the unrotated case. This suggests it will be the preferred orientation at close range, but slightly less so at larger range.

We choose the axis of rotation to be the $x_1$-axis. The sense of the $90^\circ$ rotation does not matter due to the symmetry of the solution. Using the same procedure as for the unrotated case but with
\begin{equation}
R^{\color{DarkBlue}{(2)}} = \left(
\begin{matrix}
1 & 0 & 0 \\
0 & 0 & 1 \\
0 &-1 & 0
\end{matrix}
\right) \text{ ,}
\end{equation}
we find the following result for the quadrupole interaction at general separation $\vec X$:
\begin{align}
E_\text{int,rot}^{l=2}(\vec X) =& \frac{16 C_2^2}{9}\Bigg(q_4(|\vec X|)\Big((3\hat X_1^2-1)^2+2(3\hat X_2^2-1)(3\hat X_3^2-1)\Big)\nonumber\\
&+\frac{q_3(|\vec X|)}{|\vec X|}(36 \hat X_1^2-12)\Bigg) \text{ .} \label{eq:l2intr}
\end{align}
Symmetry tells us the octupole part will change its sign compared to the unrotated case, $E_\text{int,rot}^{l=3}=-E_\text{int}^{l=3}$ (eq. (\ref{eq:l3int})) and again $E_\text{int,rot}^{l\leq3}(\vec X)=E_\text{int,rot}^{l=2}(\vec X)+E_\text{int,rot}^{l=3}(\vec X)$. As previously, a formula in terms of familiar functions can be obtained in the massless pion case:
\begin{align}
E_\text{int,rot}^{l\leq3}\Bigg|_{m=0} =& \frac{12C_2^2}{\pi} \frac{1}{|\vec X|^5} \left(-35\hat X_1^4-70\hat X_2^2 \hat X_3^2+20\hat X_1^2+5\right)\nonumber\\
&- \frac{15 C_3^2}{\pi} \frac{1}{|\vec X|^7} \left( 462 \hat X_1^2 \hat X_2^2 \hat X_3^2  + 21 \sum_{c=1}^{3} \hat X_c^4 - 17 \right) \text{ .}
\end{align}

\section{Numerical computation of the constants $C_2$ and $C_3$}

We have not yet determined the asymptotic coefficients $C_2$ and $C_3$, as they are not accessible to any but the crudest analytic approximations. To find $C_2$ and $C_3$, we need to solve a three-dimensional PDE, whereas the hedgehog asymptotic $C_1$ could be found using ODE methods (see section \ref{sec:c1det}).

\subsection{Method}

A major problem with the computation of asymptotic values is the finite box size: Computing a solution of the Skyrme model in a finite box with vacuum $U=1$ boundary conditions cuts off the asymptotics at the limits of the box. Worse, it has a large and hard-to-estimate influence even inside the box. Therefore, we choose a different approach: as the form of the asymptotic decay is known, we can impose this as a boundary condition, choosing some trial value for the constants $C_2, C_3$. Then the energy inside the box $E_\text{inside}(C_2,C_3)$ can be computed using full three-dimensional numerics (by nonlinear PDE techniques), whereas the small energy outside the box $E_\text{outside}(C_2,C_3)$ can be estimated using the leading asymptotics by means of an integration.

The constants are then varied and the minimum of the total energy $E=E_\text{inside}(C_2,C_3)+E_\text{outside}(C_2,C_3)$ is determined. This will give their numerical values.

To compute the energy inside the box of size $L$, we discretize the Skyrme energy density using central 6th order finite differences on a cubic lattice with lattice spacing $h=L/N$, where $N+1$ is the number of grid points in each direction. This gives the energy as a function of a discrete field. A nonlinear conjugate gradient method\footnote{Precisely the ``Nonlinear Conjugate Gradients with Secant and Polak-Ribière'' method described in \cite{shewchuk1994} on page 53, without preconditioning.} is then employed to find the minimum of this energy. The main program was implemented in the C programming language with thread-level parallelisation for computing gradients.

To increase consistency, $E_\text{outside}$ is determined using the same discretisation scheme as inside. The $E_\text{outside}$ integral is computed in a box of total side length $16$ in the $m=1.0$ case, $24$ in the $m=0$ case considered below. These side lengths are chosen so that the neglected energy is lower than the error bound for the extrapolated total energy. We are first treating the $m=1.0$ case interesting for nuclear physics. 

\subsection{Numerical results}

\begin{figure}[!bp]
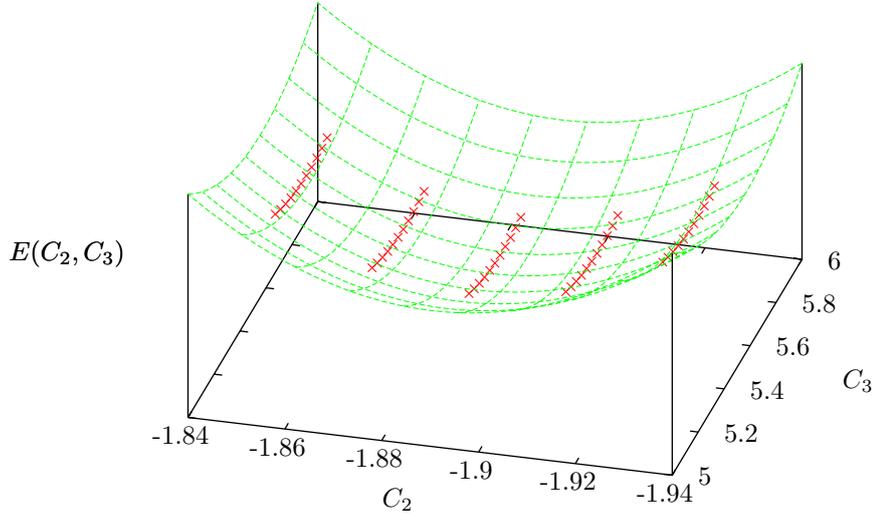

        \centering
        \input minimum-m1-90-0_08.tex
        \caption{The energy $E$ for a Skyrme field inside a finite box can be computed for a range of values of $C_2$ and $C_3$. Taking the value of $C_2$ and $C_3$ where $E$ is minimal leads to the static solution with the correct asymptotics. The crosses represent computed values, the mesh is a quadratic fit. This graph is for $h=0.08$, $L=7.20$.}
        \label{fig:minimum}
\end{figure}

\begin{table} \centering
\begin{tabular}{rr|rrrr}
$L$ & $1/L$ & $h=0.2$ & $h=0.16$ & $h=0.1$ & $h=0.08$ \\
\hline
4.8 & 0.208 & -1.783 & -1.779 & -1.766 & -1.761\\
5.6 & 0.179 & -1.841 &        & -1.838 & -1.836\\
6.4 & 0.156 & -1.870 & -1.873 & -1.871 & -1.870\\
7.2 & 0.139 & -1.888 &        & -1.891 & -1.890\\
\end{tabular}
\caption{Numerical value of $C_2$ (quadrupole constant) depending on $L$ and $h$ at $m=1.0$}
\label{tab:C2}
\vspace{5ex}
\begin{tabular}{rr|rrrr}
$L$ & $1/L$ & $h=0.2$ & $h=0.16$ & $h=0.1$ & $h=0.08$ \\
\hline
4.8 & 0.208 & 5.391 & 5.389 & 5.361 & 5.350 \\
5.6 & 0.179 & 5.495 &       & 5.501 & 5.498\\
6.4 & 0.156 & 5.531 & 5.546 & 5.546 & 5.545\\
7.2 & 0.139 & 5.548 &       & 5.566 & 5.566\\
\end{tabular}
\caption{Numerical value of $C_3$ (octupole constant) depending on $L$ and $h$ at $m=1.0$}
\label{tab:C3}
\vspace{5ex}
\begin{tabular}{rr|rrrr} 
$L$ & $1/L$ & $h=0.2$ & $h=0.16$ & $h=0.1$ & $h=0.08$ \\
\hline
4.8 & 0.208 & 5.1794 & 5.1808 & 5.1815 & 5.1816\\
5.6 & 0.179 & 5.1786 &        & 5.1802 & 5.1802\\
6.4 & 0.156 & 5.1784 & 5.1796 & 5.1800 & 5.1800\\
7.2 & 0.139 & 5.1784 &        & 5.1799 & 5.1800\\
\end{tabular}
\caption{Numerical value of $E$ (the total energy of the $B=4$ Skyrmion) depending on $L$ and $h$ at $m=1.0$}
\label{tab:E4}
\end{table}

It is expected that both the lattice spacing, $h$, and the box size $L=N\cdot h$ have an influence on the accuracy of the computations. Given infinite numerical precision, the limit of $h\rightarrow0$ and $L\rightarrow\infty$ should lead to convergence. Numerical precision and computation time put constraints on $N$ however, which limits both $L$ and $h$. In order to have a better control over the influence of $h$ and $L$ on the constants $C_2$ and $C_3$, we vary $h$ and $L$ determining the constants $C_2$ and $C_3$ in each case.

For given $h$ and $L$, $E$ is computed for a grid of values of $C_2$ and $C_3$, leading to a graph as displayed in figure \ref{fig:minimum}. A quadratic is fitted to these results to minimise $E$ and arrive at the best estimate for $C_2$ and $C_3$ at this $h$ and $L$.

The different values are shown in tables \ref{tab:C2} and \ref{tab:C3}.

\begin{figure}[!t]
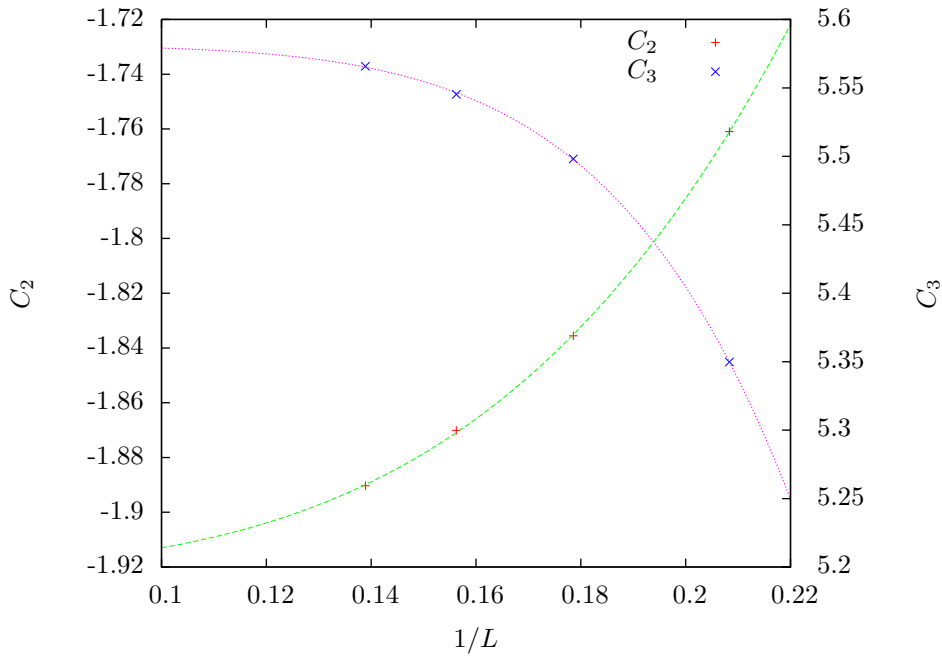
 
        \centering
        \input extraconst.tex
        \caption{$C_2$ and $C_3$ for different values of $1/L$. The value at $1/L\rightarrow0$ is determined by fitting a $A+B\cdot(1/L)^C$ function; $A$ is then the extrapolated value.}
        \label{fig:extraconst}
\end{figure}

It can be seen from these tables that the influence of the lattice spacing $h$ on the resultant parameters is rather small, whereas the box size $L$ has a significant influence. To extract the final values, we take the $h=0.08$ results and extrapolate the values to $1/L\rightarrow0$ using an $A+B\cdot(1/L)^C$ ansatz (see figure \ref{fig:extraconst}). Fitting these, we find
\begin{align}
C_2|_{m=1.0}= \ -&1.921\pm0.004\nonumber\\
C_3|_{m=1.0}= \ &5.581 \pm0.003 \text{ .}
\end{align}
Additionally, this technique gives us the possibility to get a value of the energy of the $B=4$ Skyrmion at infinite box size. The same extrapolation technique applied to table \ref{tab:E4} gives the value 
\begin{equation}
E|_{m=1.0}=5.1799\pm0.0002\text{ .}
\end{equation}
The errors given are taken from the fit errors of the extrapolation of $1/L\rightarrow0$, except for the energy, where the error from computing at finite $h$ is dominant and is estimated using the results in table \ref{tab:E4}.

The $B=4$ energy at $m=1$ has been computed before as $E\approx5.23$ in \cite{battye2006} and $E=5.27$ in \cite{battye2009}. While the deviation from our value seems significant, only a precision of about $\unit[0.5]{\%}$ was claimed there.

\begin{table}[bp!]
\begin{minipage}{0.45\linewidth}\centering
\begin{tabular}{rr|rr}
$L$ & $1/L$ & $h=0.16$ & $h=0.1$ \\
\hline
5.6 & 0.179 &        & -2.506 \\
6.4 & 0.156 &        & -2.763 \\
7.2 & 0.139 &        & -2.892 \\
8.0 & 0.125 & -2.968 & -2.961 \\
\end{tabular}
\caption{Numerical $C_2$ at $m=0$}
\label{tab:C20}
\end{minipage}
\begin{minipage}{0.45\linewidth}\centering
\begin{tabular}{rr|rr}
$L$ & $1/L$ & $h=0.16$ & $h=0.1$ \\
\hline
5.6 & 0.179 &        & 10.960 \\
6.4 & 0.156 &        & 11.952 \\
7.2 & 0.139 &        & 12.352 \\
8.0 & 0.125 & 12.530 & 12.520 \\
\end{tabular}
\caption{Numerical $C_3$ at $m=0$}
\label{tab:C30}
\end{minipage}
\centering
\begin{minipage}{0.45\linewidth}\centering
\vspace{4ex}
\begin{tabular}{rr|rr}
$L$ & $1/L$ & $h=0.16$ & $h=0.1$ \\
\hline
5.6 & 0.179 &        & 4.4904 \\
6.4 & 0.156 &        & 4.4830 \\
7.2 & 0.139 &        & 4.4809 \\
8.0 & 0.125 & 4.4801 & 4.4802 \\
\end{tabular}
\caption{Numerical $E$ at $m=0$}
\label{tab:E40}
\end{minipage}
\end{table}
The $m=0$ case is numerically somewhat more challenging, as the fields decay more slowly and the box has to be larger in order to obtain the same accuracy. From our experience in the massive case, we compute at different box sizes $L$ for extrapolation, leaving $h=0.1$ constant. Only for $L=8.0$, we also compute at $h=0.16$ in order to be able to estimate errors from the lattice spacing $h$ (see tables \ref{tab:C20}--\ref{tab:E40}). By fitting $A+B\cdot(1/L)^C$ to the results in the $h=0.1$ columns, we find the values
\begin{align}
C_2|_{m=0}              &= -3.077\pm0.010 \nonumber\\
C_3|_{m=0}              &= 12.706\pm0.020\\
E|_{m=0}               &= 4.4798\pm0.0002\text{ .} \nonumber
\end{align}
The energy agrees with the value given in \cite[p. 377]{manton2004}. In this ($m=0$) case, the fit error of the $1/L\rightarrow0$ extrapolation is much smaller for both $C_2$ and $E$, so that the error estimates above are the approximate variations between the $h=0.16$ and the $h=0.1$ values. For $C_3$, both errors have a similar magnitude, so twice the fit error is given as the error estimate.

\section{Application to $B=8$, $B=12$ and the Skyrme crystal}

\subsection{The $B=8$ Skyrmion}
\begin{figure}[!bp]
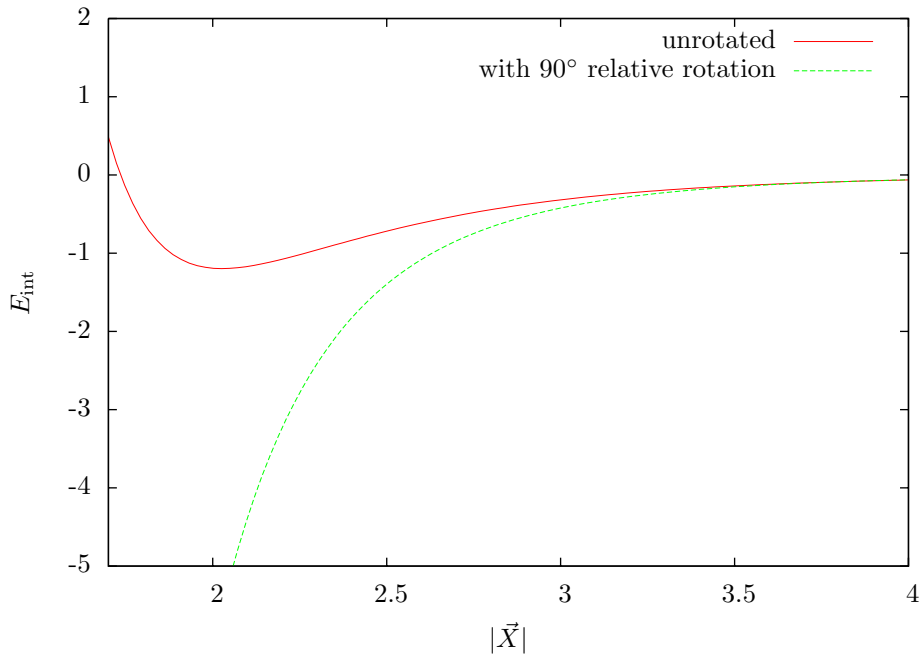

        \centering
        \input eintmrur.tex
        \caption{Interaction energies $E_\text{int}^{l\leq3}$ for two $B=4$ Skyrmions separated along the $x_1$-axis ($m=1.0$); red/solid in the same orientation, green/dashed with $90^\circ$ relative rotation around the $x_1$-axis. While the asymptotic interaction energy in the rotated case goes to $-\infty$ as $|\vec X|\rightarrow 0$, there will actually be nonlinear corrections leading to a positive interaction energy at short distances.}
        \label{fig:eintmrur}
\end{figure}

A comparison of the asymptotic interaction energies of two $B=4$ Skyrmions in the unrotated and $90^\circ$ rotated case is shown in figure \ref{fig:eintmrur}. This is for the separation vector $\vec X=(X,0,0)$. The asymptotic interaction energy in the unrotated case is slightly lower for large $|\vec X|$. However, at short distances, the rotated configuration has lower energy, due to the large contribution from the negative octupole interaction energy. The latter going to $-\infty$ at the origin (zero separation) is clearly an artifact of the asymptotic approximation and the Skyrmions will repel at close distances.

For $m=1.0$, the numerically determined $B=8$ Skyrmion consists of two $B=4$ cubes, rotated by $90^\circ$ around the axis of separation. In \cite[p. 95]{wood2009}, the distance between the cubes has been determined using the moment of inertia and found to be $|\vec X|\approx 3.0$. At this separation, our calculations show clearly that it is energetically favourable to rotate one Skyrmion by $90^\circ$, rather than leave it unrotated. The full interaction energy can be found using the energy of the $B=8$ Skyrmion and subtracting two times the energy of the $B=4$ Skyrmion (all energies taken from \cite{battye2009}) and is\footnote{This number is the difference of two numerical energies of which the precision is not known and could therefore have a significant error.} $E_\text{int}=-0.30$. Inserting $\vec X=(3.0,0,0)$ into the asymptotic interaction energy, we predict a value of $E_\text{int,rot}^{l\leq3}(\vec X)=-0.42$. This is surprisingly close to the full interaction energy given that the separation is relatively small.

\subsection{The linear $B=12$ configuration}

One of the candidates for a minimal energy solution for $B=12$ is a row of three $B=4$ Skyrmions, where the outer ones are in the same orientation while the middle one is rotated by $90^\circ$ around their axis. Placing the three $B=4$ cubes at $-\vec X = (-d,0,0)$, $(0,0,0)$ and $\vec X = (d,0,0)$, their asymptotic interaction energy is given by
\begin{equation}
E_\text{int,lin12}^{l\leq3} = 2E_\text{int,rot}^{l\leq3}(\vec X)+E_\text{int}^{l\leq3}(2\vec X)
\end{equation}
where the interaction energies are as defined in sections \ref{sec:eint} and \ref{sec:eintr}. Due to the exponential decay, the last term has a negligible influence, leading us to predict that $d=3.0$ as in the $B=8$ case (the same assumption was made in \cite[p. 106]{wood2009}). Here the predicted interaction energy is $E_\text{int,lin12}^{l\leq3}=-0.85$. Unfortunately, a numerical energy of the linear $B=12$ solution of sufficient accuracy does not seem to be available at this point, so we cannot compare this prediction to any value in the literature.

\subsection{The Skyrme crystal}

\begin{figure}[tbp]
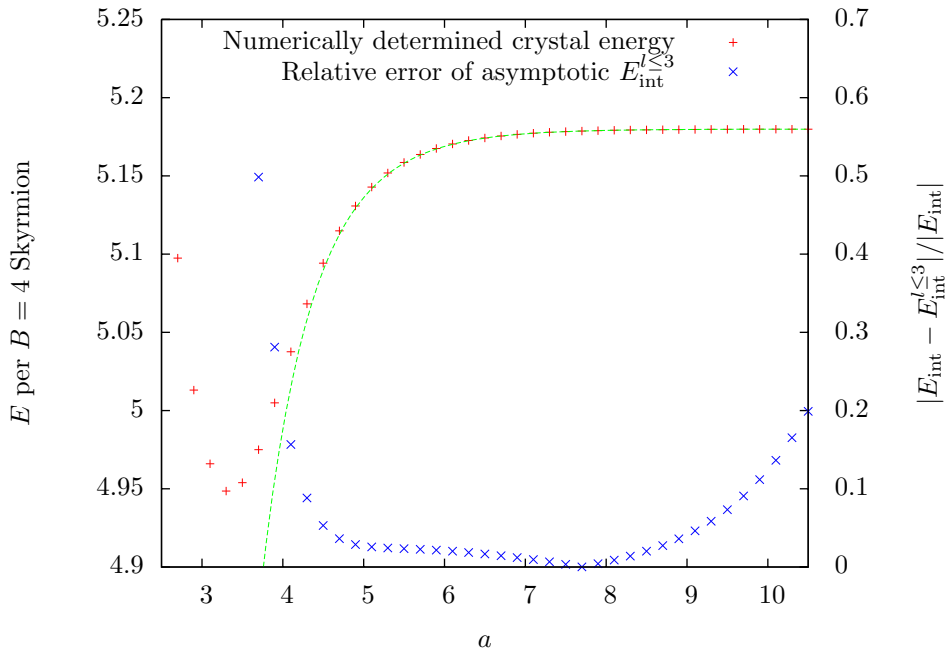

        \centering
        \input crystalem.tex
        \caption{The energy $E$ per $B=4$ Skyrmion for the Skyrme crystal at $m=1.0$ as function of the lattice constant $a$. The green/dashed line is the prediction from asymptotic quadrupole and octupole interactions. The relative error of the asymptotic approximation is also displayed. Note that for $a\gtrsim 7.5$ the relative error is mainly due to the error in the total energy, suggesting the asymptotic interaction energy is actually much more precise.}
        \label{fig:skyrmecrystm}
\end{figure}
\begin{figure}[tbp]
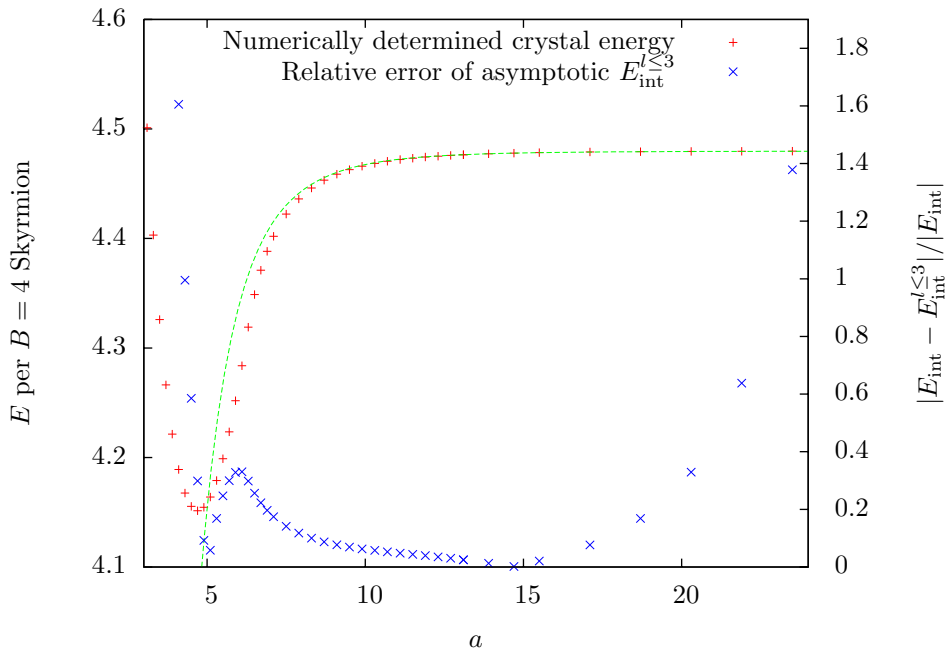

        \centering
        \input crystale.tex
        \caption{The energy $E$ per $B=4$ Skyrmion for the Skyrme crystal at $m=0$ as function of the lattice constant $a$. The green/dashed line is the prediction from asymptotic quadrupole and octupole interactions. The relative error for $a\gtrsim 15$ is, similar to the $m=1.0$ case, dominated by the error in determining $E$. For $m=0$, we expect a larger relative error also due to contributions from more distant lattice sites as well as the interaction generally decaying more slowly.}
        \label{fig:skyrmecryst}
\end{figure}
The Skyrme crystal can be thought of as an infinite simple cubic (sc) lattice of $B=4$ Skyrmions, all in the same orientation (compare \cite{kugler1988}). Although we have found that at close range, a relative rotation of $90^\circ$ will give a stronger face-to-face attraction, it is not possible to construct a crystal with all cubes in face contact rotated by $90^\circ$ relatively around their separation vector.

If we let $a$ be the smallest distance between two $B=4$ Skyrmions, then for any $a$ we can compute the asymptotic interaction energy of an infinite lattice of these. Summing the nearest-neighbour contributions from the $B=4$ Skyrmions adjacent through the 6 faces, the 12 edges and the 8 corners to a given $B=4$ Skyrmion, and dividing by 2, this asymptotic interaction is
\begin{align}
E_\text{int,crystal}^{l\leq3}\Bigg|_{m=0} =& \frac{7 \left(128 \sqrt{3}-3888+243 \sqrt{2}\right) C_2^2}{54 \pi } \frac{1}{a^5}\nonumber\\
&+\frac{5 \left(4096\sqrt{3}-139968 +28431 \sqrt{2}\right) C_3^2}{3888 \pi } \frac{1}{a^7}
\end{align}
per $B=4$ Skyrmion for $m=0$. In the massive pion case, the interaction energy can also be determined, and is illustrated below, but results in a far more complicated formula.

The true energy $E$ per unit cell in the crystal is rather easy to find numerically, as it just corresponds to finding a solution with $B=4$ whilst imposing periodic boundary conditions. The box size $L$ is then the same as the distance $a$ between two neighbouring $B=4$ Skyrmions. This has been done for the massive pion (figure \ref{fig:skyrmecrystm}) and massless (figure \ref{fig:skyrmecryst}) case. Additionally the predicted value from the asymptotic interaction is displayed.

It can be seen that for $m=1.0$, the asymptotic interaction energy predicts the crystal energy very well for lattice constants greater then about $a=4$. The error however starts to grow again for large $a$. This is because it is numerically impossible to directly determine $E_\text{int}$. To determine $E_\text{int}$, we have to compute the energy of the free $B=4$ Skyrmion and subtract it from the energy in the unit cell of the crystal, which gives the determined $E_\text{int}$ a large relative error.

For the $m=0$ case, the prediction is not as good. This can be traced back to the lower precisions of the constants $C_2$, $C_3$ and $E$ in this case (due to the higher computational complexity of determining them) and non-negligible contributions from more distant Skyrmions in the interactions. Also, due to the slower decay, it is expected that this interaction is only valid at much larger range than in the massive pion case.

While in both cases, the octupole and the quadrupole parts of the interactions feature different signs and the predicted asymptotic interaction therefore has a minimum (compare figure \ref{fig:eintmrur}), this is unfortunately clearly not in a range where the interactions can be trusted at all. Therefore, the asymptotic interactions could not be used to predict the Skyrme crystal lattice constant on their own: For $m=0$, the lattice constant we find is $a\approx4.71$ with an energy of $E\approx4.15$ per unit cell. The values found previously are $a\approx4.7$ \cite{kugler1988}\footnote{Notice that due to their definition of the Skyrme Lagrangian, length units are $1/(e F_\pi)$ instead of $2/(e F_\pi)$ used here. However, their $L$ is the distance between two half-Skyrmions, and not two $B=4$ configurations as is our $a$ here. So coincidentally the values of $a$ here and $L$ in \cite{kugler1988} are numerically the same.} with an energy of $E\approx4.15$ per $B=4$ unit cell. In the $m=1.0$ case, the Skyrmion crystal was not computed before. From the numerical values presented in figure \ref{fig:skyrmecrystm} here, we find 
\begin{align}
a|_{m=1.0}&\approx3.35 \nonumber\\
E|_{m=1.0}&\approx4.95 \text{ per $B=4$ Skyrmion}
\end{align}
using a quadratic interpolation through the three values closest to the minimum. This can be compared to the energy value found in \cite{lee2003} at $m\approx0.32$, which is\footnote{Their $L$ is given in the same units as used here, but $a=2L$.} $E\approx4.3$ at $a\approx4.4$.

\begin{figure}[bp!]\centering
\includegraphics[width=0.45\textwidth]{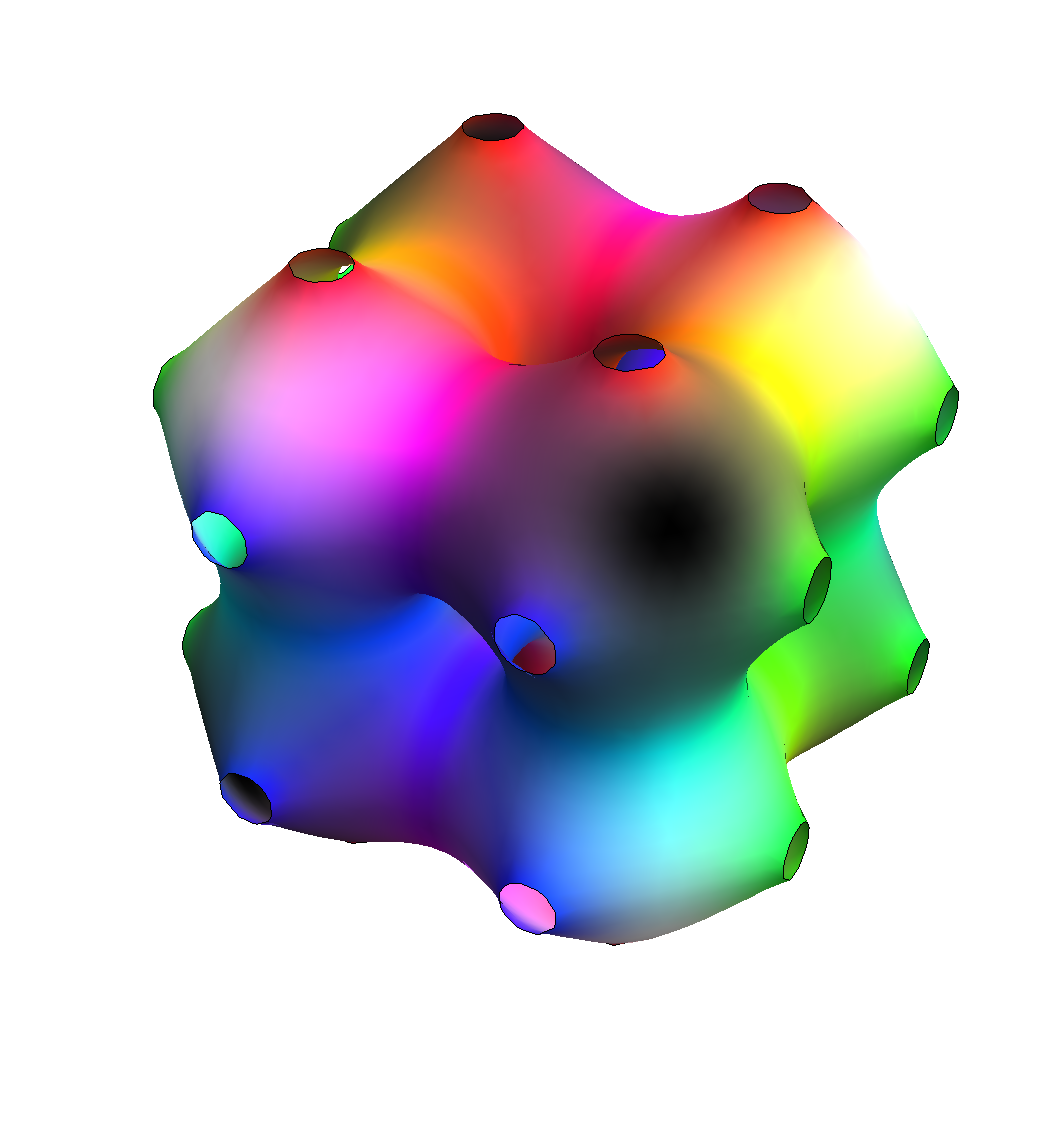}
\includegraphics[width=0.45\textwidth]{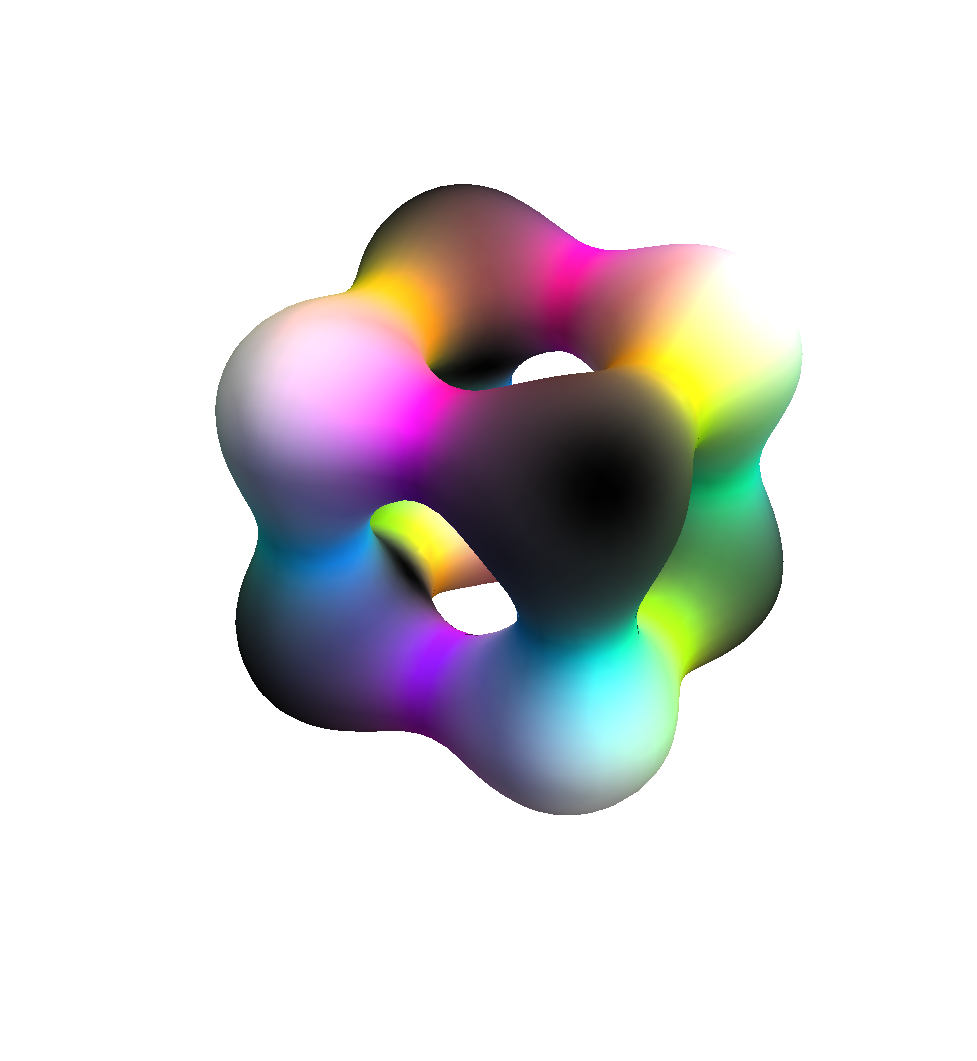}
\caption{Two isobaryon (left: $B^0=0.1$, right: $B^0=0.2$) surfaces of the $m=1.0$ Skyrme crystal at the minimum ($a=3.3$). It can be seen that a unit cell contains a cube that is more strongly connected internally than to the next unit cell.}
\label{fig:crystcont}
\end{figure}
In figure \ref{fig:crystcont}, two isobaryon surfaces of the $m=1.0$ Skyrme crystal are displayed. Computing at the lattice spacing $h=0.1$ restricts us to box sizes of $a=(N+1)h$, so we chose $a=3.30$ which is very close to the minimum.

For the $m=0$ Skyrme crystal, a translation by $a/2$ in one of the axis directions leaves the baryon and energy density invariant. Our isobaryon densities show that this is not the case for the $m=1.0$ Skyrmion crystal, which is due to the asymmetry between $U=1$ and $U=-1$ created by the mass term (see eq. (\ref{eq:skyrme})). In our configuration, $U=-1$ at the centre of the cube and $U=1$ at the boundaries of the unit cell.

\section{Conclusion}

In this paper, we have shown in detail how to find the asymptotic interactions of the $B=4$ Skyrmions, and laid out a general framework which can easily be applied to other solutions of the Skyrme model or soliton interactions in general, with massless or massive pions.

The $B=4$ solution is very important in the Skyrme model, as it appears as a building block of many higher baryon number solutions. Working out its interactions can lead to a better understanding of how these $B=4$ blocks arrange themselves in a more complex solution. With this work, the interactions have now been understood both qualitatively, by putting them into a nice computational framework using the three-colour scheme, and quantitatively, by determining the constants $C_2$ and $C_3$. This has been applied to the $B=8$ and $B=12$ Skyrmions to estimate their interaction energies. A range of validity for the asymptotic interaction energy has been established using the Skyrme crystal. Together with $C_1$, the hedgehog dipole constant, $C_2$ and $C_3$ encode our knowledge about the interactions of the most important solutions of the Skyrme model in the context of nuclear physics.

Further research can now build on this framework to see how much the asymptotic interactions can serve to predict the form of solutions of the Skyrme model. It has been shown before that this approach works well in the case of four $B=1$ hedgehogs making up a $B=4$ solution, where the asymptotic interaction predicts how they should be oriented \cite{manton1994}. If the same could be done for solutions built from $B=4$ blocks, or a mixture of $B=4$ blocks and hedgehogs, this would constitute a tremendous help in the search for solutions.

While the value of $C_1$ on its own would have limited relevance to a qualitative structural understanding, as that is determined by the angular dependence of the interaction, the case of $B=4$ is different (because there are two constants, so their relative strengths are important). In the case of systems consisting of $B=1$ and $B=4$ blocks, all three constants will come into play -- and they could be interpreted as the structure constants of the Skyrme model, determining the composition rules of larger Skyrmions.

\section{Acknowledgements}

I thank the following people for their support in this work: Nick Manton, my supervisor, for regular help and discussions, Juha Jäykkä for pointing out the correct discretisation procedure for the Skyrme model, Thomas Fischbacher, helping with improving the numerics and Guido Franchetti for helping with visualisations.

My work is supported by the Gates Cambridge Trust and EPSRC.

\addcontentsline{toc}{section}{References}
\bibliographystyle{amsunsrt}
\bibliography{df-skinter}

\end{document}

%% file: massparams10.tex
\begingroup
  \makeatletter
  \providecommand\color[2][]{%
    \GenericError{(gnuplot) \space\space\space\@spaces}{%
      Package color not loaded in conjunction with
      terminal option `colourtext'%
    }{See the gnuplot documentation for explanation.%
    }{Either use 'blacktext' in gnuplot or load the package
      color.sty in LaTeX.}%
    \renewcommand\color[2][]{}%
  }%
  \providecommand\includegraphics[2][]{%
    \GenericError{(gnuplot) \space\space\space\@spaces}{%
      Package graphicx or graphics not loaded%
    }{See the gnuplot documentation for explanation.%
    }{The gnuplot epslatex terminal needs graphicx.sty or graphics.sty.}%
    \renewcommand\includegraphics[2][]{}%
  }%
  \providecommand\rotatebox[2]{#2}%
  \@ifundefined{ifGPcolor}{%
    \newif\ifGPcolor
    \GPcolortrue
  }{}%
  \@ifundefined{ifGPblacktext}{%
    \newif\ifGPblacktext
    \GPblacktextfalse
  }{}%
  \let\gplgaddtomacro\g@addto@macro
  \gdef\gplbacktext{}%
  \gdef\gplfronttext{}%
  \makeatother
  \ifGPblacktext
    \def\colorrgb#1{}%
    \def\colorgray#1{}%
  \else
    \ifGPcolor
      \def\colorrgb#1{\color[rgb]{#1}}%
      \def\colorgray#1{\color[gray]{#1}}%
      \expandafter\def\csname LTw\endcsname{\color{white}}%
      \expandafter\def\csname LTb\endcsname{\color{black}}%
      \expandafter\def\csname LTa\endcsname{\color{black}}%
      \expandafter\def\csname LT0\endcsname{\color[rgb]{1,0,0}}%
      \expandafter\def\csname LT1\endcsname{\color[rgb]{0,1,0}}%
      \expandafter\def\csname LT2\endcsname{\color[rgb]{0,0,1}}%
      \expandafter\def\csname LT3\endcsname{\color[rgb]{1,0,1}}%
      \expandafter\def\csname LT4\endcsname{\color[rgb]{0,1,1}}%
      \expandafter\def\csname LT5\endcsname{\color[rgb]{1,1,0}}%
      \expandafter\def\csname LT6\endcsname{\color[rgb]{0,0,0}}%
      \expandafter\def\csname LT7\endcsname{\color[rgb]{1,0.3,0}}%
      \expandafter\def\csname LT8\endcsname{\color[rgb]{0.5,0.5,0.5}}%
    \else
      \def\colorrgb#1{\color{black}}%
      \def\colorgray#1{\color[gray]{#1}}%
      \expandafter\def\csname LTw\endcsname{\color{white}}%
      \expandafter\def\csname LTb\endcsname{\color{black}}%
      \expandafter\def\csname LTa\endcsname{\color{black}}%
      \expandafter\def\csname LT0\endcsname{\color{black}}%
      \expandafter\def\csname LT1\endcsname{\color{black}}%
      \expandafter\def\csname LT2\endcsname{\color{black}}%
      \expandafter\def\csname LT3\endcsname{\color{black}}%
      \expandafter\def\csname LT4\endcsname{\color{black}}%
      \expandafter\def\csname LT5\endcsname{\color{black}}%
      \expandafter\def\csname LT6\endcsname{\color{black}}%
      \expandafter\def\csname LT7\endcsname{\color{black}}%
      \expandafter\def\csname LT8\endcsname{\color{black}}%
    \fi
  \fi
  \setlength{\unitlength}{0.0500bp}%
  \begin{picture}(7200.00,5040.00)%
    \gplgaddtomacro\gplbacktext{%
      \csname LTb\endcsname%
      \put(726,660){\makebox(0,0)[r]{\strut{} 0}}%
      \put(726,1346){\makebox(0,0)[r]{\strut{} 1}}%
      \put(726,2032){\makebox(0,0)[r]{\strut{} 2}}%
      \put(726,2718){\makebox(0,0)[r]{\strut{} 3}}%
      \put(726,3404){\makebox(0,0)[r]{\strut{} 4}}%
      \put(726,4090){\makebox(0,0)[r]{\strut{} 5}}%
      \put(726,4776){\makebox(0,0)[r]{\strut{} 6}}%
      \put(858,440){\makebox(0,0){\strut{} 0}}%
      \put(2052,440){\makebox(0,0){\strut{} 2}}%
      \put(3245,440){\makebox(0,0){\strut{} 4}}%
      \put(4439,440){\makebox(0,0){\strut{} 6}}%
      \put(5632,440){\makebox(0,0){\strut{} 8}}%
      \put(6826,440){\makebox(0,0){\strut{} 10}}%
      \put(220,2718){\rotatebox{90}{\makebox(0,0){\strut{}$C_1$}}}%
      \put(3842,110){\makebox(0,0){\strut{}$m$}}%
    }%
    \gplgaddtomacro\gplfronttext{%
    }%
    \gplbacktext
    \put(0,0){\includegraphics{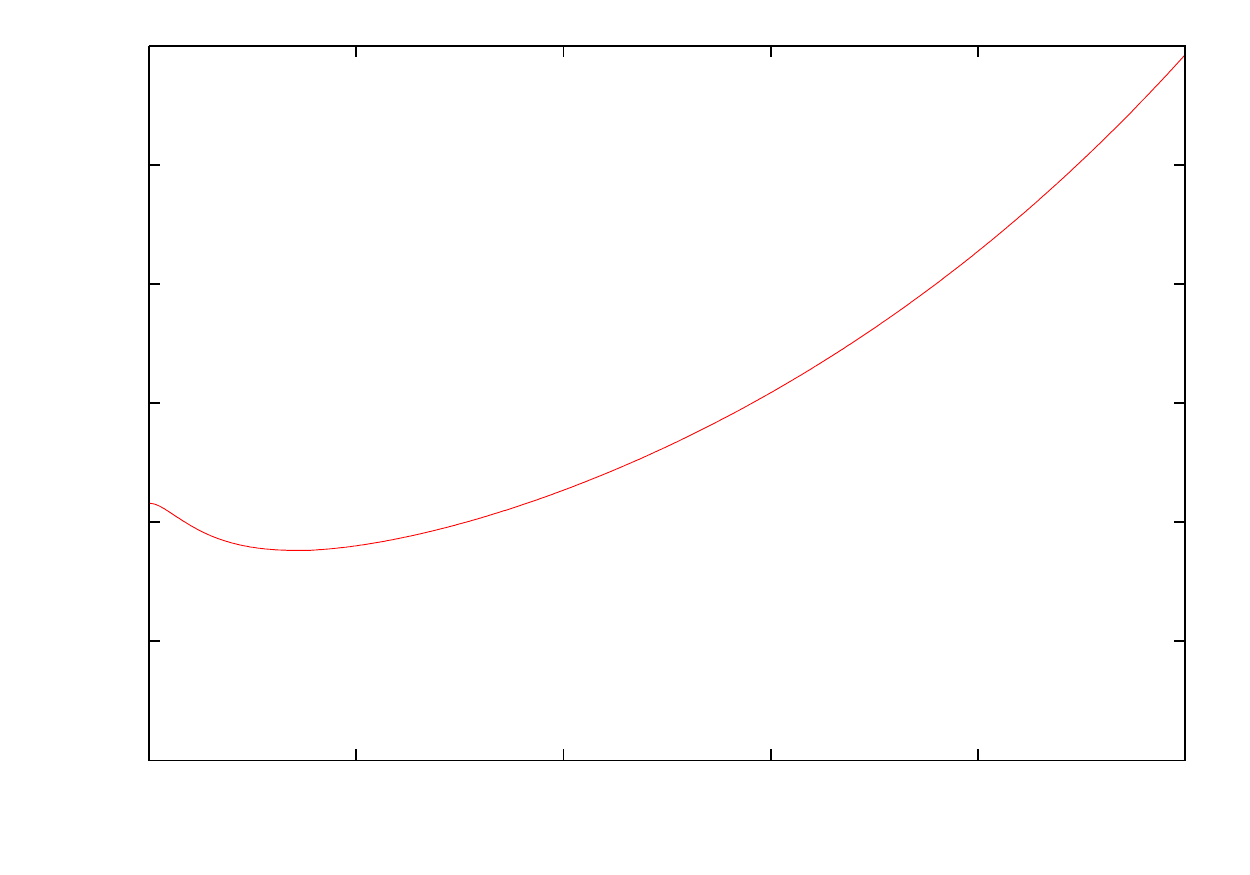}}%
    \gplfronttext
  \end{picture}%
\endgroup

%% file: genericmultisetting.pdftex_t
\begin{picture}(0,0)%
\includegraphics{genericmultisetting.pdftex}%
\end{picture}%
\setlength{\unitlength}{1381sp}%
\begingroup\makeatletter\ifx\SetFigFont\undefined%
\gdef\SetFigFont#1#2#3#4#5{%
  \reset@font\fontsize{#1}{#2pt}%
  \fontfamily{#3}\fontseries{#4}\fontshape{#5}%
  \selectfont}%
\fi\endgroup%
\begin{picture}(13773,6036)(-3674,-5164)
\put(4576,-3136){\makebox(0,0)[lb]{\smash{{\SetFigFont{14}{16.8}{\rmdefault}{\mddefault}{\updefault}{\color[rgb]{0,0,0}$\vec X$}%
}}}}
\put(-1499,-3961){\makebox(0,0)[lb]{\smash{{\SetFigFont{14}{16.8}{\rmdefault}{\mddefault}{\updefault}{\color[rgb]{0,0,0}$\rho^{\color{DarkRed}{(1)}}$}%
}}}}
\put(9151,-1561){\makebox(0,0)[lb]{\smash{{\SetFigFont{14}{16.8}{\rmdefault}{\mddefault}{\updefault}{\color[rgb]{0,0,0}$\rho^{\color{DarkBlue}{(2)}}$}%
}}}}
\put(-974,464){\makebox(0,0)[lb]{\smash{{\SetFigFont{14}{16.8}{\rmdefault}{\mddefault}{\updefault}{\color[rgb]{0,0,0}$x_2$}%
}}}}
\put(-3674,-5011){\makebox(0,0)[lb]{\smash{{\SetFigFont{14}{16.8}{\rmdefault}{\mddefault}{\updefault}{\color[rgb]{0,0,0}$x_3$}%
}}}}
\put(2776,-3511){\makebox(0,0)[lb]{\smash{{\SetFigFont{14}{16.8}{\rmdefault}{\mddefault}{\updefault}{\color[rgb]{0,0,0}$x_1$}%
}}}}
\end{picture}%

%% file: coord_1.pdftex_t
\begin{picture}(0,0)%
\includegraphics{coord_1.pdftex}%
\end{picture}%
\setlength{\unitlength}{987sp}%
\begingroup\makeatletter\ifx\SetFigFont\undefined%
\gdef\SetFigFont#1#2#3#4#5{%
  \reset@font\fontsize{#1}{#2pt}%
  \fontfamily{#3}\fontseries{#4}\fontshape{#5}%
  \selectfont}%
\fi\endgroup%
\begin{picture}(3052,3411)(3737,-5089)
\put(5251,-4936){\makebox(0,0)[lb]{\smash{{\SetFigFont{10}{12.0}{\rmdefault}{\mddefault}{\updefault}{\color[rgb]{0,0,0}$x_1$}%
}}}}
\put(4426,-3436){\makebox(0,0)[lb]{\smash{{\SetFigFont{10}{12.0}{\rmdefault}{\mddefault}{\updefault}{\color[rgb]{0,0,0}$x_2$}%
}}}}
\put(3976,-2086){\makebox(0,0)[lb]{\smash{{\SetFigFont{10}{12.0}{\rmdefault}{\mddefault}{\updefault}{\color[rgb]{0,0,0}$x_3$}%
}}}}
\end{picture}%

%% file: coord_1pi.pdftex_t
\begin{picture}(0,0)%
\includegraphics{coord_1pi.pdftex}%
\end{picture}%
\setlength{\unitlength}{987sp}%
\begingroup\makeatletter\ifx\SetFigFont\undefined%
\gdef\SetFigFont#1#2#3#4#5{%
  \reset@font\fontsize{#1}{#2pt}%
  \fontfamily{#3}\fontseries{#4}\fontshape{#5}%
  \selectfont}%
\fi\endgroup%
\begin{picture}(1933,3477)(3737,-5107)
\put(4426,-3436){\makebox(0,0)[lb]{\smash{{\SetFigFont{10}{12.0}{\rmdefault}{\mddefault}{\updefault}{\color[rgb]{0,0,0}$\pi_2$}%
}}}}
\put(3976,-2086){\makebox(0,0)[lb]{\smash{{\SetFigFont{10}{12.0}{\rmdefault}{\mddefault}{\updefault}{\color[rgb]{0,0,0}$\pi_3$}%
}}}}
\put(5251,-4936){\makebox(0,0)[lb]{\smash{{\SetFigFont{10}{12.0}{\rmdefault}{\mddefault}{\updefault}{\color[rgb]{0,0,0}$\pi_1$}%
}}}}
\end{picture}%

%% file: coord_2.pdftex_t
\begin{picture}(0,0)%
\includegraphics{coord_2.pdftex}%
\end{picture}%
\setlength{\unitlength}{1776sp}%
\begingroup\makeatletter\ifx\SetFigFont\undefined%
\gdef\SetFigFont#1#2#3#4#5{%
  \reset@font\fontsize{#1}{#2pt}%
  \fontfamily{#3}\fontseries{#4}\fontshape{#5}%
  \selectfont}%
\fi\endgroup%
\begin{picture}(3000,3402)(3301,-5032)
\put(6301,-4411){\makebox(0,0)[lb]{\smash{{\SetFigFont{17}{20.4}{\rmdefault}{\mddefault}{\updefault}{\color[rgb]{0,0,0}$\pi_1$}%
}}}}
\put(3976,-2086){\makebox(0,0)[lb]{\smash{{\SetFigFont{17}{20.4}{\rmdefault}{\mddefault}{\updefault}{\color[rgb]{0,0,0}$\pi_2$}%
}}}}
\put(3301,-4861){\makebox(0,0)[lb]{\smash{{\SetFigFont{17}{20.4}{\rmdefault}{\mddefault}{\updefault}{\color[rgb]{0,0,0}$\pi_3$}%
}}}}
\end{picture}%

%% file: alphaqvectors.pdftex_t
\begin{picture}(0,0)%
\includegraphics{alphaqvectors.pdftex}%
\end{picture}%
\setlength{\unitlength}{1776sp}%
\begingroup\makeatletter\ifx\SetFigFont\undefined%
\gdef\SetFigFont#1#2#3#4#5{%
  \reset@font\fontsize{#1}{#2pt}%
  \fontfamily{#3}\fontseries{#4}\fontshape{#5}%
  \selectfont}%
\fi\endgroup%
\begin{picture}(3375,4134)(2251,-4861)
\put(3526,-1111){\makebox(0,0)[lb]{\smash{{\SetFigFont{17}{20.4}{\rmdefault}{\mddefault}{\updefault}{\color[rgb]{0,0,0}$\hat x^3$}%
}}}}
\put(5626,-3886){\makebox(0,0)[lb]{\smash{{\SetFigFont{17}{20.4}{\rmdefault}{\mddefault}{\updefault}{\color[rgb]{0,0,0}$\hat x^1$}%
}}}}
\put(2776,-4636){\makebox(0,0)[lb]{\smash{{\SetFigFont{17}{20.4}{\rmdefault}{\mddefault}{\updefault}{\color[rgb]{0,0,0}$\hat x^2$}%
}}}}
\end{picture}%

%% file: minimum-m1-90-0_08.tex
\begingroup
  \makeatletter
  \providecommand\color[2][]{%
    \GenericError{(gnuplot) \space\space\space\@spaces}{%
      Package color not loaded in conjunction with
      terminal option `colourtext'%
    }{See the gnuplot documentation for explanation.%
    }{Either use 'blacktext' in gnuplot or load the package
      color.sty in LaTeX.}%
    \renewcommand\color[2][]{}%
  }%
  \providecommand\includegraphics[2][]{%
    \GenericError{(gnuplot) \space\space\space\@spaces}{%
      Package graphicx or graphics not loaded%
    }{See the gnuplot documentation for explanation.%
    }{The gnuplot epslatex terminal needs graphicx.sty or graphics.sty.}%
    \renewcommand\includegraphics[2][]{}%
  }%
  \providecommand\rotatebox[2]{#2}%
  \@ifundefined{ifGPcolor}{%
    \newif\ifGPcolor
    \GPcolortrue
  }{}%
  \@ifundefined{ifGPblacktext}{%
    \newif\ifGPblacktext
    \GPblacktexttrue
  }{}%
  \let\gplgaddtomacro\g@addto@macro
  \gdef\gplbacktext{}%
  \gdef\gplfronttext{}%
  \makeatother
  \ifGPblacktext
    \def\colorrgb#1{}%
    \def\colorgray#1{}%
  \else
    \ifGPcolor
      \def\colorrgb#1{\color[rgb]{#1}}%
      \def\colorgray#1{\color[gray]{#1}}%
      \expandafter\def\csname LTw\endcsname{\color{white}}%
      \expandafter\def\csname LTb\endcsname{\color{black}}%
      \expandafter\def\csname LTa\endcsname{\color{black}}%
      \expandafter\def\csname LT0\endcsname{\color[rgb]{1,0,0}}%
      \expandafter\def\csname LT1\endcsname{\color[rgb]{0,1,0}}%
      \expandafter\def\csname LT2\endcsname{\color[rgb]{0,0,1}}%
      \expandafter\def\csname LT3\endcsname{\color[rgb]{1,0,1}}%
      \expandafter\def\csname LT4\endcsname{\color[rgb]{0,1,1}}%
      \expandafter\def\csname LT5\endcsname{\color[rgb]{1,1,0}}%
      \expandafter\def\csname LT6\endcsname{\color[rgb]{0,0,0}}%
      \expandafter\def\csname LT7\endcsname{\color[rgb]{1,0.3,0}}%
      \expandafter\def\csname LT8\endcsname{\color[rgb]{0.5,0.5,0.5}}%
    \else
      \def\colorrgb#1{\color{black}}%
      \def\colorgray#1{\color[gray]{#1}}%
      \expandafter\def\csname LTw\endcsname{\color{white}}%
      \expandafter\def\csname LTb\endcsname{\color{black}}%
      \expandafter\def\csname LTa\endcsname{\color{black}}%
      \expandafter\def\csname LT0\endcsname{\color{black}}%
      \expandafter\def\csname LT1\endcsname{\color{black}}%
      \expandafter\def\csname LT2\endcsname{\color{black}}%
      \expandafter\def\csname LT3\endcsname{\color{black}}%
      \expandafter\def\csname LT4\endcsname{\color{black}}%
      \expandafter\def\csname LT5\endcsname{\color{black}}%
      \expandafter\def\csname LT6\endcsname{\color{black}}%
      \expandafter\def\csname LT7\endcsname{\color{black}}%
      \expandafter\def\csname LT8\endcsname{\color{black}}%
    \fi
  \fi
  \setlength{\unitlength}{0.0500bp}%
  \begin{picture}(7200.00,5040.00)%
    \gplgaddtomacro\gplbacktext{%
      \csname LTb\endcsname%
      \put(4868,611){\makebox(0,0){\strut{}-1.94}}%
      \put(4146,698){\makebox(0,0){\strut{}-1.92}}%
      \put(3424,785){\makebox(0,0){\strut{}-1.9}}%
      \put(2701,872){\makebox(0,0){\strut{}-1.88}}%
      \put(1979,959){\makebox(0,0){\strut{}-1.86}}%
      \put(1256,1046){\makebox(0,0){\strut{}-1.84}}%
      \put(5049,736){\makebox(0,0)[l]{\strut{} 5}}%
      \put(5243,1061){\makebox(0,0)[l]{\strut{} 5.2}}%
      \put(5436,1385){\makebox(0,0)[l]{\strut{} 5.4}}%
      \put(5630,1710){\makebox(0,0)[l]{\strut{} 5.6}}%
      \put(5824,2034){\makebox(0,0)[l]{\strut{} 5.8}}%
      \put(6017,2359){\makebox(0,0)[l]{\strut{} 6}}%
      \put(407,2425){\makebox(0,0){\strut{}$E(C_2,C_3)$}}%
    }%
    \gplgaddtomacro\gplfronttext{%
      \csname LTb\endcsname%
      \put(2875,573){\makebox(0,0){\strut{}$C_2$}}%
      \put(6309,1464){\makebox(0,0){\strut{}$C_3$}}%
      \put(407,2425){\makebox(0,0){\strut{}$E(C_2,C_3)$}}%
    }%
    \gplbacktext
    \put(0,0){\includegraphics{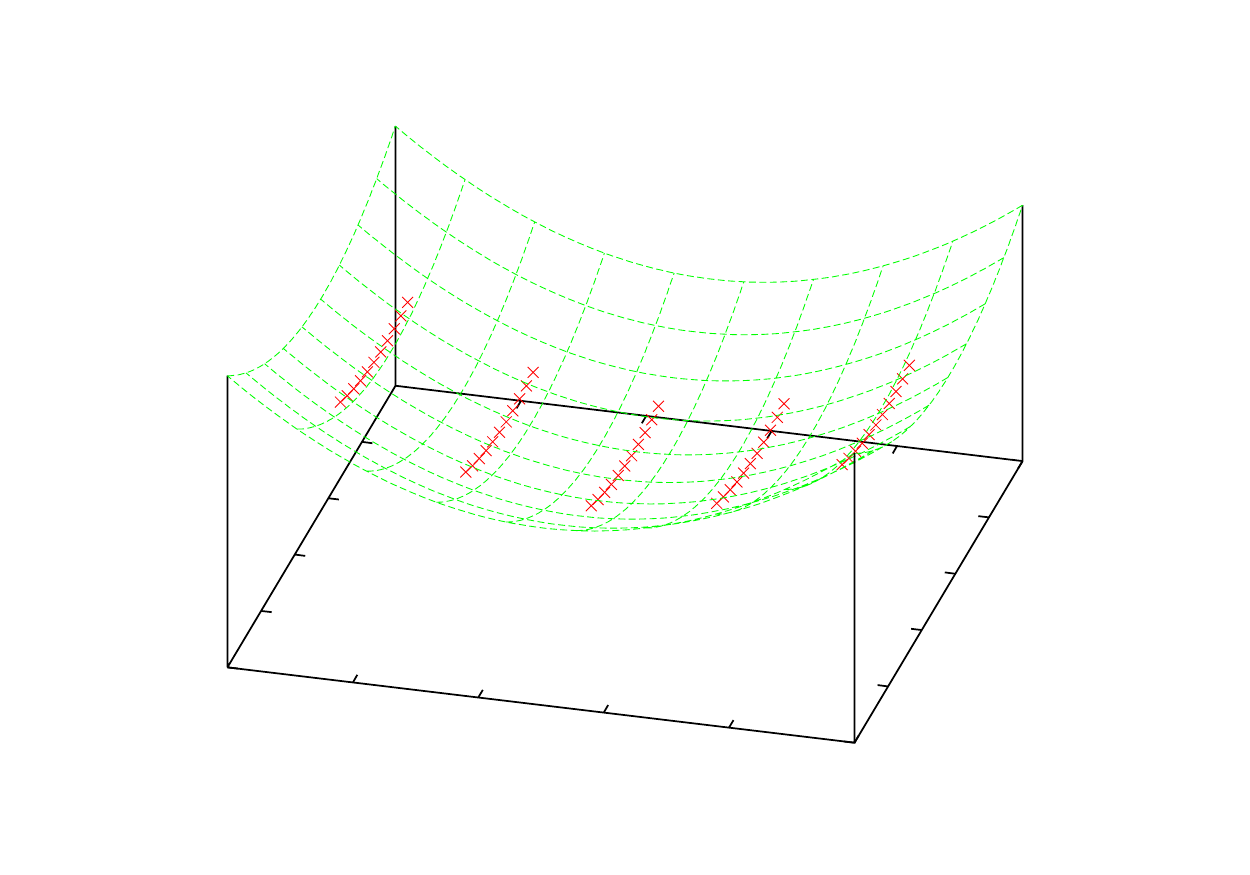}}%
    \gplfronttext
  \end{picture}%
\endgroup

%% file: extraconst.tex
\begingroup
  \makeatletter
  \providecommand\color[2][]{%
    \GenericError{(gnuplot) \space\space\space\@spaces}{%
      Package color not loaded in conjunction with
      terminal option `colourtext'%
    }{See the gnuplot documentation for explanation.%
    }{Either use 'blacktext' in gnuplot or load the package
      color.sty in LaTeX.}%
    \renewcommand\color[2][]{}%
  }%
  \providecommand\includegraphics[2][]{%
    \GenericError{(gnuplot) \space\space\space\@spaces}{%
      Package graphicx or graphics not loaded%
    }{See the gnuplot documentation for explanation.%
    }{The gnuplot epslatex terminal needs graphicx.sty or graphics.sty.}%
    \renewcommand\includegraphics[2][]{}%
  }%
  \providecommand\rotatebox[2]{#2}%
  \@ifundefined{ifGPcolor}{%
    \newif\ifGPcolor
    \GPcolortrue
  }{}%
  \@ifundefined{ifGPblacktext}{%
    \newif\ifGPblacktext
    \GPblacktexttrue
  }{}%
  \let\gplgaddtomacro\g@addto@macro
  \gdef\gplbacktext{}%
  \gdef\gplfronttext{}%
  \makeatother
  \ifGPblacktext
    \def\colorrgb#1{}%
    \def\colorgray#1{}%
  \else
    \ifGPcolor
      \def\colorrgb#1{\color[rgb]{#1}}%
      \def\colorgray#1{\color[gray]{#1}}%
      \expandafter\def\csname LTw\endcsname{\color{white}}%
      \expandafter\def\csname LTb\endcsname{\color{black}}%
      \expandafter\def\csname LTa\endcsname{\color{black}}%
      \expandafter\def\csname LT0\endcsname{\color[rgb]{1,0,0}}%
      \expandafter\def\csname LT1\endcsname{\color[rgb]{0,1,0}}%
      \expandafter\def\csname LT2\endcsname{\color[rgb]{0,0,1}}%
      \expandafter\def\csname LT3\endcsname{\color[rgb]{1,0,1}}%
      \expandafter\def\csname LT4\endcsname{\color[rgb]{0,1,1}}%
      \expandafter\def\csname LT5\endcsname{\color[rgb]{1,1,0}}%
      \expandafter\def\csname LT6\endcsname{\color[rgb]{0,0,0}}%
      \expandafter\def\csname LT7\endcsname{\color[rgb]{1,0.3,0}}%
      \expandafter\def\csname LT8\endcsname{\color[rgb]{0.5,0.5,0.5}}%
    \else
      \def\colorrgb#1{\color{black}}%
      \def\colorgray#1{\color[gray]{#1}}%
      \expandafter\def\csname LTw\endcsname{\color{white}}%
      \expandafter\def\csname LTb\endcsname{\color{black}}%
      \expandafter\def\csname LTa\endcsname{\color{black}}%
      \expandafter\def\csname LT0\endcsname{\color{black}}%
      \expandafter\def\csname LT1\endcsname{\color{black}}%
      \expandafter\def\csname LT2\endcsname{\color{black}}%
      \expandafter\def\csname LT3\endcsname{\color{black}}%
      \expandafter\def\csname LT4\endcsname{\color{black}}%
      \expandafter\def\csname LT5\endcsname{\color{black}}%
      \expandafter\def\csname LT6\endcsname{\color{black}}%
      \expandafter\def\csname LT7\endcsname{\color{black}}%
      \expandafter\def\csname LT8\endcsname{\color{black}}%
    \fi
  \fi
  \setlength{\unitlength}{0.0500bp}%
  \begin{picture}(7200.00,5040.00)%
    \gplgaddtomacro\gplbacktext{%
      \csname LTb\endcsname%
      \put(1122,660){\makebox(0,0)[r]{\strut{}-1.92}}%
      \put(1122,1072){\makebox(0,0)[r]{\strut{}-1.9}}%
      \put(1122,1483){\makebox(0,0)[r]{\strut{}-1.88}}%
      \put(1122,1895){\makebox(0,0)[r]{\strut{}-1.86}}%
      \put(1122,2306){\makebox(0,0)[r]{\strut{}-1.84}}%
      \put(1122,2718){\makebox(0,0)[r]{\strut{}-1.82}}%
      \put(1122,3130){\makebox(0,0)[r]{\strut{}-1.8}}%
      \put(1122,3541){\makebox(0,0)[r]{\strut{}-1.78}}%
      \put(1122,3953){\makebox(0,0)[r]{\strut{}-1.76}}%
      \put(1122,4364){\makebox(0,0)[r]{\strut{}-1.74}}%
      \put(1122,4776){\makebox(0,0)[r]{\strut{}-1.72}}%
      \put(1254,440){\makebox(0,0){\strut{} 0.1}}%
      \put(2036,440){\makebox(0,0){\strut{} 0.12}}%
      \put(2818,440){\makebox(0,0){\strut{} 0.14}}%
      \put(3600,440){\makebox(0,0){\strut{} 0.16}}%
      \put(4382,440){\makebox(0,0){\strut{} 0.18}}%
      \put(5164,440){\makebox(0,0){\strut{} 0.2}}%
      \put(5946,440){\makebox(0,0){\strut{} 0.22}}%
      \put(6078,660){\makebox(0,0)[l]{\strut{} 5.2}}%
      \put(6078,1174){\makebox(0,0)[l]{\strut{} 5.25}}%
      \put(6078,1689){\makebox(0,0)[l]{\strut{} 5.3}}%
      \put(6078,2203){\makebox(0,0)[l]{\strut{} 5.35}}%
      \put(6078,2718){\makebox(0,0)[l]{\strut{} 5.4}}%
      \put(6078,3232){\makebox(0,0)[l]{\strut{} 5.45}}%
      \put(6078,3747){\makebox(0,0)[l]{\strut{} 5.5}}%
      \put(6078,4261){\makebox(0,0)[l]{\strut{} 5.55}}%
      \put(6078,4776){\makebox(0,0)[l]{\strut{} 5.6}}%
      \put(220,2718){\rotatebox{90}{\makebox(0,0){\strut{}$C_2$}}}%
      \put(6979,2718){\rotatebox{90}{\makebox(0,0){\strut{}$C_3$}}}%
      \put(3600,110){\makebox(0,0){\strut{}$1/L$}}%
    }%
    \gplgaddtomacro\gplfronttext{%
      \csname LTb\endcsname%
      \put(4959,4603){\makebox(0,0)[r]{\strut{}$C_2$}}%
      \csname LTb\endcsname%
      \put(4959,4383){\makebox(0,0)[r]{\strut{}$C_3$}}%
    }%
    \gplbacktext
    \put(0,0){\includegraphics{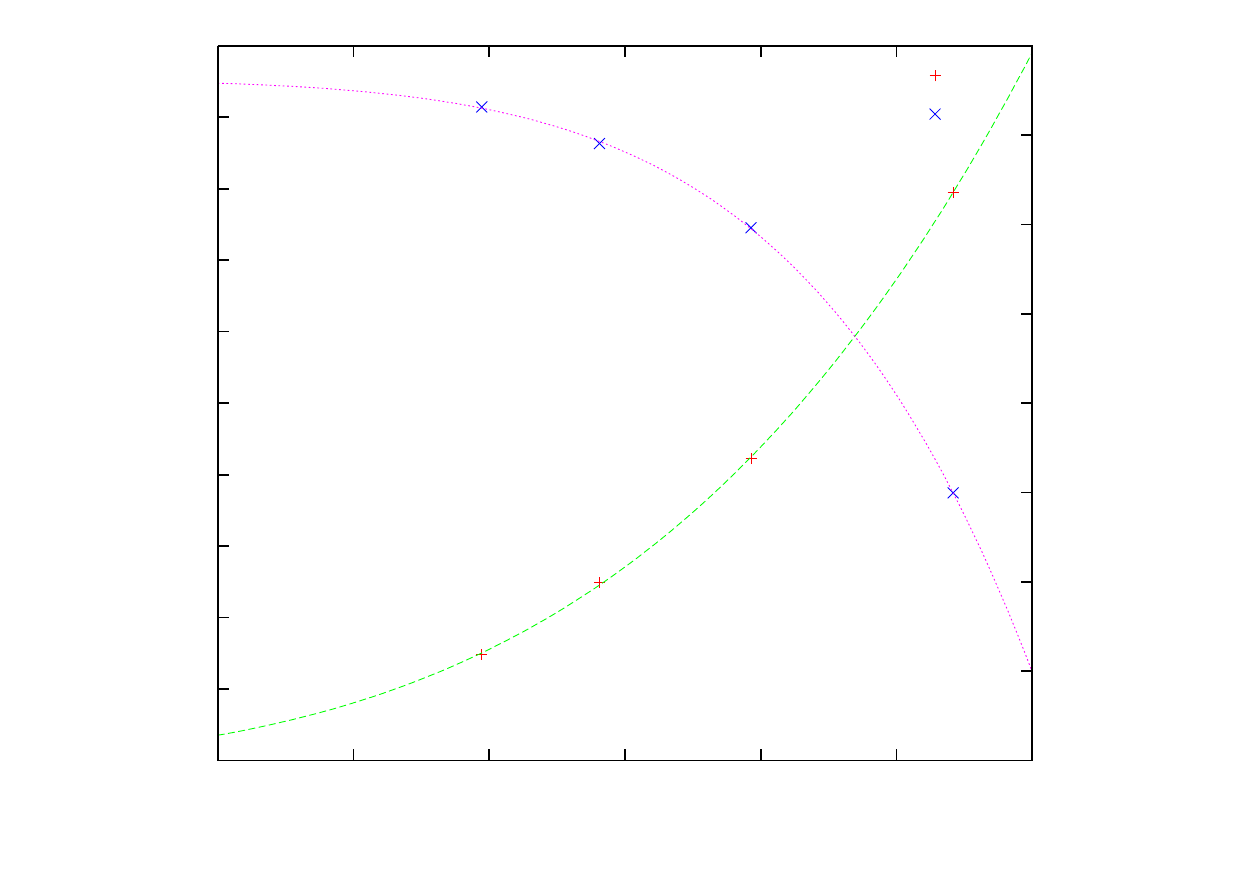}}%
    \gplfronttext
  \end{picture}%
\endgroup

%% file: eintmrur.tex
\begingroup
  \makeatletter
  \providecommand\color[2][]{%
    \GenericError{(gnuplot) \space\space\space\@spaces}{%
      Package color not loaded in conjunction with
      terminal option `colourtext'%
    }{See the gnuplot documentation for explanation.%
    }{Either use 'blacktext' in gnuplot or load the package
      color.sty in LaTeX.}%
    \renewcommand\color[2][]{}%
  }%
  \providecommand\includegraphics[2][]{%
    \GenericError{(gnuplot) \space\space\space\@spaces}{%
      Package graphicx or graphics not loaded%
    }{See the gnuplot documentation for explanation.%
    }{The gnuplot epslatex terminal needs graphicx.sty or graphics.sty.}%
    \renewcommand\includegraphics[2][]{}%
  }%
  \providecommand\rotatebox[2]{#2}%
  \@ifundefined{ifGPcolor}{%
    \newif\ifGPcolor
    \GPcolortrue
  }{}%
  \@ifundefined{ifGPblacktext}{%
    \newif\ifGPblacktext
    \GPblacktexttrue
  }{}%
  \let\gplgaddtomacro\g@addto@macro
  \gdef\gplbacktext{}%
  \gdef\gplfronttext{}%
  \makeatother
  \ifGPblacktext
    \def\colorrgb#1{}%
    \def\colorgray#1{}%
  \else
    \ifGPcolor
      \def\colorrgb#1{\color[rgb]{#1}}%
      \def\colorgray#1{\color[gray]{#1}}%
      \expandafter\def\csname LTw\endcsname{\color{white}}%
      \expandafter\def\csname LTb\endcsname{\color{black}}%
      \expandafter\def\csname LTa\endcsname{\color{black}}%
      \expandafter\def\csname LT0\endcsname{\color[rgb]{1,0,0}}%
      \expandafter\def\csname LT1\endcsname{\color[rgb]{0,1,0}}%
      \expandafter\def\csname LT2\endcsname{\color[rgb]{0,0,1}}%
      \expandafter\def\csname LT3\endcsname{\color[rgb]{1,0,1}}%
      \expandafter\def\csname LT4\endcsname{\color[rgb]{0,1,1}}%
      \expandafter\def\csname LT5\endcsname{\color[rgb]{1,1,0}}%
      \expandafter\def\csname LT6\endcsname{\color[rgb]{0,0,0}}%
      \expandafter\def\csname LT7\endcsname{\color[rgb]{1,0.3,0}}%
      \expandafter\def\csname LT8\endcsname{\color[rgb]{0.5,0.5,0.5}}%
    \else
      \def\colorrgb#1{\color{black}}%
      \def\colorgray#1{\color[gray]{#1}}%
      \expandafter\def\csname LTw\endcsname{\color{white}}%
      \expandafter\def\csname LTb\endcsname{\color{black}}%
      \expandafter\def\csname LTa\endcsname{\color{black}}%
      \expandafter\def\csname LT0\endcsname{\color{black}}%
      \expandafter\def\csname LT1\endcsname{\color{black}}%
      \expandafter\def\csname LT2\endcsname{\color{black}}%
      \expandafter\def\csname LT3\endcsname{\color{black}}%
      \expandafter\def\csname LT4\endcsname{\color{black}}%
      \expandafter\def\csname LT5\endcsname{\color{black}}%
      \expandafter\def\csname LT6\endcsname{\color{black}}%
      \expandafter\def\csname LT7\endcsname{\color{black}}%
      \expandafter\def\csname LT8\endcsname{\color{black}}%
    \fi
  \fi
  \setlength{\unitlength}{0.0500bp}%
  \begin{picture}(7200.00,5040.00)%
    \gplgaddtomacro\gplbacktext{%
      \csname LTb\endcsname%
      \put(726,660){\makebox(0,0)[r]{\strut{}-5}}%
      \put(726,1248){\makebox(0,0)[r]{\strut{}-4}}%
      \put(726,1836){\makebox(0,0)[r]{\strut{}-3}}%
      \put(726,2424){\makebox(0,0)[r]{\strut{}-2}}%
      \put(726,3012){\makebox(0,0)[r]{\strut{}-1}}%
      \put(726,3600){\makebox(0,0)[r]{\strut{} 0}}%
      \put(726,4188){\makebox(0,0)[r]{\strut{} 1}}%
      \put(726,4776){\makebox(0,0)[r]{\strut{} 2}}%
      \put(1636,440){\makebox(0,0){\strut{} 2}}%
      \put(2934,440){\makebox(0,0){\strut{} 2.5}}%
      \put(4231,440){\makebox(0,0){\strut{} 3}}%
      \put(5529,440){\makebox(0,0){\strut{} 3.5}}%
      \put(6826,440){\makebox(0,0){\strut{} 4}}%
      \put(220,2718){\rotatebox{90}{\makebox(0,0){\strut{}$E_\text{int}$}}}%
      \put(3842,110){\makebox(0,0){\strut{}$|\vec X|$}}%
    }%
    \gplgaddtomacro\gplfronttext{%
      \csname LTb\endcsname%
      \put(5839,4603){\makebox(0,0)[r]{\strut{}unrotated}}%
      \csname LTb\endcsname%
      \put(5839,4383){\makebox(0,0)[r]{\strut{}with $90^\circ$ relative rotation}}%
    }%
    \gplbacktext
    \put(0,0){\includegraphics{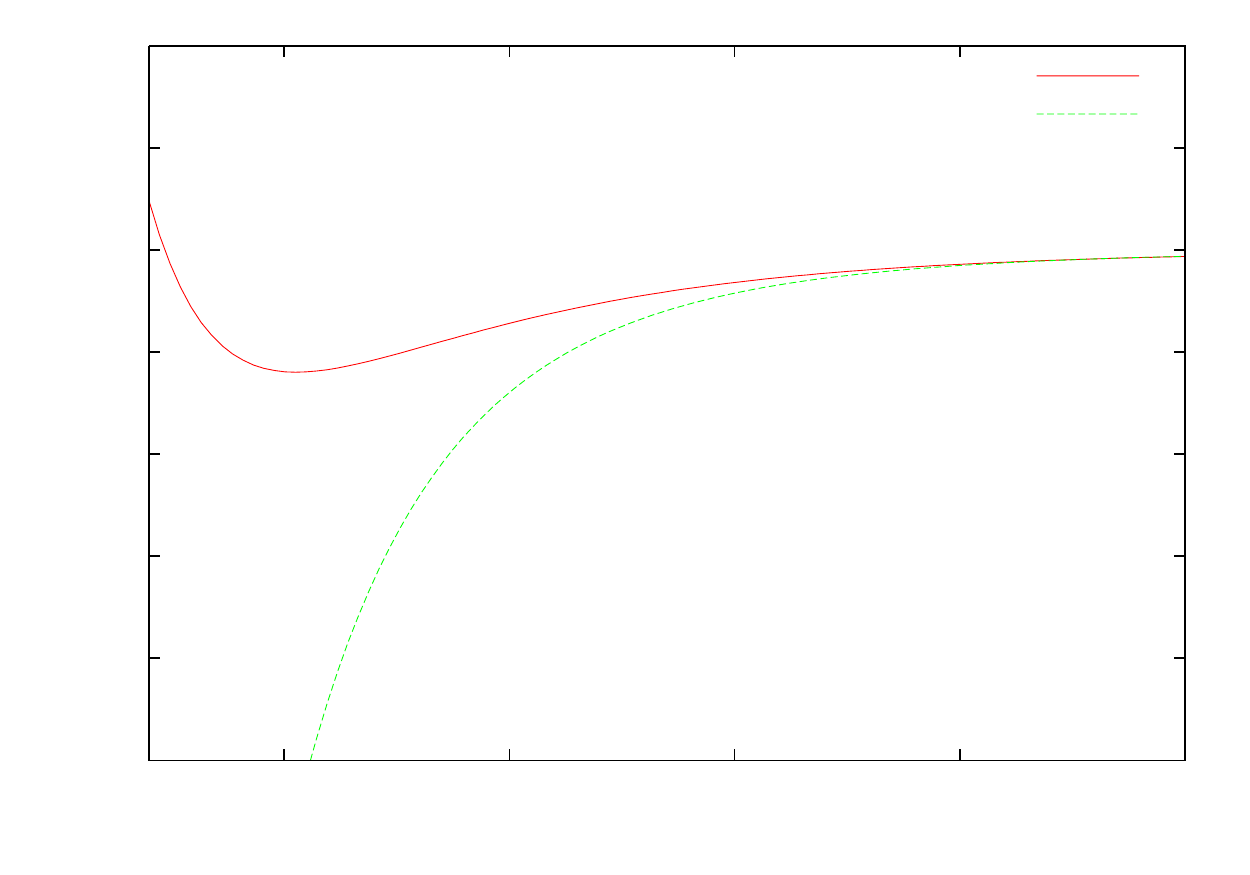}}%
    \gplfronttext
  \end{picture}%
\endgroup

%% file: crystalem.tex
\begingroup
  \makeatletter
  \providecommand\color[2][]{%
    \GenericError{(gnuplot) \space\space\space\@spaces}{%
      Package color not loaded in conjunction with
      terminal option `colourtext'%
    }{See the gnuplot documentation for explanation.%
    }{Either use 'blacktext' in gnuplot or load the package
      color.sty in LaTeX.}%
    \renewcommand\color[2][]{}%
  }%
  \providecommand\includegraphics[2][]{%
    \GenericError{(gnuplot) \space\space\space\@spaces}{%
      Package graphicx or graphics not loaded%
    }{See the gnuplot documentation for explanation.%
    }{The gnuplot epslatex terminal needs graphicx.sty or graphics.sty.}%
    \renewcommand\includegraphics[2][]{}%
  }%
  \providecommand\rotatebox[2]{#2}%
  \@ifundefined{ifGPcolor}{%
    \newif\ifGPcolor
    \GPcolortrue
  }{}%
  \@ifundefined{ifGPblacktext}{%
    \newif\ifGPblacktext
    \GPblacktexttrue
  }{}%
  \let\gplgaddtomacro\g@addto@macro
  \gdef\gplbacktext{}%
  \gdef\gplfronttext{}%
  \makeatother
  \ifGPblacktext
    \def\colorrgb#1{}%
    \def\colorgray#1{}%
  \else
    \ifGPcolor
      \def\colorrgb#1{\color[rgb]{#1}}%
      \def\colorgray#1{\color[gray]{#1}}%
      \expandafter\def\csname LTw\endcsname{\color{white}}%
      \expandafter\def\csname LTb\endcsname{\color{black}}%
      \expandafter\def\csname LTa\endcsname{\color{black}}%
      \expandafter\def\csname LT0\endcsname{\color[rgb]{1,0,0}}%
      \expandafter\def\csname LT1\endcsname{\color[rgb]{0,1,0}}%
      \expandafter\def\csname LT2\endcsname{\color[rgb]{0,0,1}}%
      \expandafter\def\csname LT3\endcsname{\color[rgb]{1,0,1}}%
      \expandafter\def\csname LT4\endcsname{\color[rgb]{0,1,1}}%
      \expandafter\def\csname LT5\endcsname{\color[rgb]{1,1,0}}%
      \expandafter\def\csname LT6\endcsname{\color[rgb]{0,0,0}}%
      \expandafter\def\csname LT7\endcsname{\color[rgb]{1,0.3,0}}%
      \expandafter\def\csname LT8\endcsname{\color[rgb]{0.5,0.5,0.5}}%
    \else
      \def\colorrgb#1{\color{black}}%
      \def\colorgray#1{\color[gray]{#1}}%
      \expandafter\def\csname LTw\endcsname{\color{white}}%
      \expandafter\def\csname LTb\endcsname{\color{black}}%
      \expandafter\def\csname LTa\endcsname{\color{black}}%
      \expandafter\def\csname LT0\endcsname{\color{black}}%
      \expandafter\def\csname LT1\endcsname{\color{black}}%
      \expandafter\def\csname LT2\endcsname{\color{black}}%
      \expandafter\def\csname LT3\endcsname{\color{black}}%
      \expandafter\def\csname LT4\endcsname{\color{black}}%
      \expandafter\def\csname LT5\endcsname{\color{black}}%
      \expandafter\def\csname LT6\endcsname{\color{black}}%
      \expandafter\def\csname LT7\endcsname{\color{black}}%
      \expandafter\def\csname LT8\endcsname{\color{black}}%
    \fi
  \fi
  \setlength{\unitlength}{0.0500bp}%
  \begin{picture}(7200.00,5040.00)%
    \gplgaddtomacro\gplbacktext{%
      \csname LTb\endcsname%
      \put(1122,660){\makebox(0,0)[r]{\strut{} 4.9}}%
      \put(1122,1248){\makebox(0,0)[r]{\strut{} 4.95}}%
      \put(1122,1836){\makebox(0,0)[r]{\strut{} 5}}%
      \put(1122,2424){\makebox(0,0)[r]{\strut{} 5.05}}%
      \put(1122,3012){\makebox(0,0)[r]{\strut{} 5.1}}%
      \put(1122,3600){\makebox(0,0)[r]{\strut{} 5.15}}%
      \put(1122,4188){\makebox(0,0)[r]{\strut{} 5.2}}%
      \put(1122,4776){\makebox(0,0)[r]{\strut{} 5.25}}%
      \put(1556,440){\makebox(0,0){\strut{} 3}}%
      \put(2159,440){\makebox(0,0){\strut{} 4}}%
      \put(2762,440){\makebox(0,0){\strut{} 5}}%
      \put(3365,440){\makebox(0,0){\strut{} 6}}%
      \put(3968,440){\makebox(0,0){\strut{} 7}}%
      \put(4571,440){\makebox(0,0){\strut{} 8}}%
      \put(5174,440){\makebox(0,0){\strut{} 9}}%
      \put(5777,440){\makebox(0,0){\strut{} 10}}%
      \put(6210,660){\makebox(0,0)[l]{\strut{} 0}}%
      \put(6210,1248){\makebox(0,0)[l]{\strut{} 0.1}}%
      \put(6210,1836){\makebox(0,0)[l]{\strut{} 0.2}}%
      \put(6210,2424){\makebox(0,0)[l]{\strut{} 0.3}}%
      \put(6210,3012){\makebox(0,0)[l]{\strut{} 0.4}}%
      \put(6210,3600){\makebox(0,0)[l]{\strut{} 0.5}}%
      \put(6210,4188){\makebox(0,0)[l]{\strut{} 0.6}}%
      \put(6210,4776){\makebox(0,0)[l]{\strut{} 0.7}}%
      \put(220,2718){\rotatebox{90}{\makebox(0,0){\strut{}$E$ per $B=4$ Skyrmion}}}%
      \put(6979,2718){\rotatebox{90}{\makebox(0,0){\strut{}$|E_\text{int}-E_\text{int}^{l\leq3}|/|E_\text{int}|$}}}%
      \put(3666,110){\makebox(0,0){\strut{}$a$}}%
    }%
    \gplgaddtomacro\gplfronttext{%
      \csname LTb\endcsname%
      \put(5091,4603){\makebox(0,0)[r]{\strut{}Numerically determined crystal energy}}%
      \csname LTb\endcsname%
      \put(5091,4383){\makebox(0,0)[r]{\strut{}Relative error of asymptotic $E_\text{int}^{l\leq3}$}}%
    }%
    \gplbacktext
    \put(0,0){\includegraphics{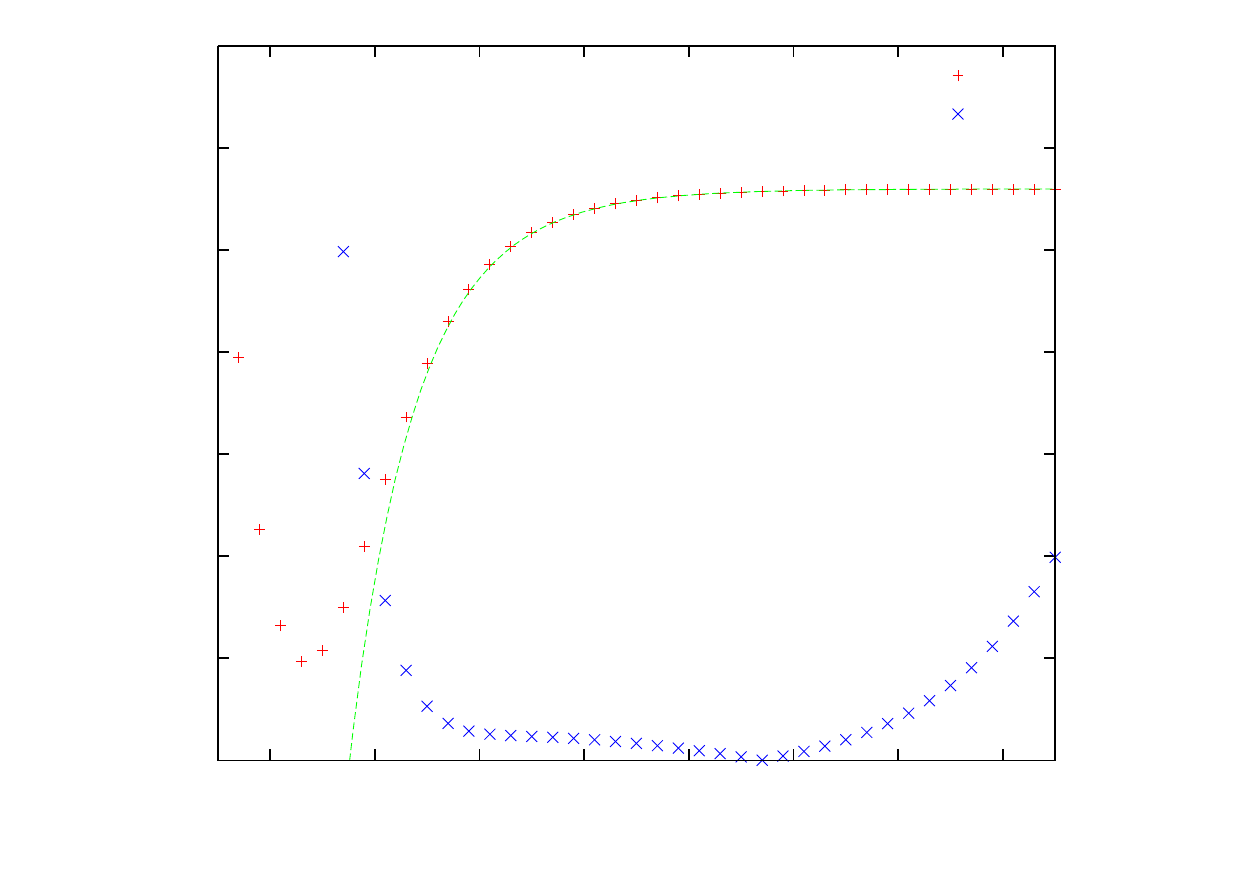}}%
    \gplfronttext
  \end{picture}%
\endgroup

%% file: crystale.tex
\begingroup
  \makeatletter
  \providecommand\color[2][]{%
    \GenericError{(gnuplot) \space\space\space\@spaces}{%
      Package color not loaded in conjunction with
      terminal option `colourtext'%
    }{See the gnuplot documentation for explanation.%
    }{Either use 'blacktext' in gnuplot or load the package
      color.sty in LaTeX.}%
    \renewcommand\color[2][]{}%
  }%
  \providecommand\includegraphics[2][]{%
    \GenericError{(gnuplot) \space\space\space\@spaces}{%
      Package graphicx or graphics not loaded%
    }{See the gnuplot documentation for explanation.%
    }{The gnuplot epslatex terminal needs graphicx.sty or graphics.sty.}%
    \renewcommand\includegraphics[2][]{}%
  }%
  \providecommand\rotatebox[2]{#2}%
  \@ifundefined{ifGPcolor}{%
    \newif\ifGPcolor
    \GPcolortrue
  }{}%
  \@ifundefined{ifGPblacktext}{%
    \newif\ifGPblacktext
    \GPblacktexttrue
  }{}%
  \let\gplgaddtomacro\g@addto@macro
  \gdef\gplbacktext{}%
  \gdef\gplfronttext{}%
  \makeatother
  \ifGPblacktext
    \def\colorrgb#1{}%
    \def\colorgray#1{}%
  \else
    \ifGPcolor
      \def\colorrgb#1{\color[rgb]{#1}}%
      \def\colorgray#1{\color[gray]{#1}}%
      \expandafter\def\csname LTw\endcsname{\color{white}}%
      \expandafter\def\csname LTb\endcsname{\color{black}}%
      \expandafter\def\csname LTa\endcsname{\color{black}}%
      \expandafter\def\csname LT0\endcsname{\color[rgb]{1,0,0}}%
      \expandafter\def\csname LT1\endcsname{\color[rgb]{0,1,0}}%
      \expandafter\def\csname LT2\endcsname{\color[rgb]{0,0,1}}%
      \expandafter\def\csname LT3\endcsname{\color[rgb]{1,0,1}}%
      \expandafter\def\csname LT4\endcsname{\color[rgb]{0,1,1}}%
      \expandafter\def\csname LT5\endcsname{\color[rgb]{1,1,0}}%
      \expandafter\def\csname LT6\endcsname{\color[rgb]{0,0,0}}%
      \expandafter\def\csname LT7\endcsname{\color[rgb]{1,0.3,0}}%
      \expandafter\def\csname LT8\endcsname{\color[rgb]{0.5,0.5,0.5}}%
    \else
      \def\colorrgb#1{\color{black}}%
      \def\colorgray#1{\color[gray]{#1}}%
      \expandafter\def\csname LTw\endcsname{\color{white}}%
      \expandafter\def\csname LTb\endcsname{\color{black}}%
      \expandafter\def\csname LTa\endcsname{\color{black}}%
      \expandafter\def\csname LT0\endcsname{\color{black}}%
      \expandafter\def\csname LT1\endcsname{\color{black}}%
      \expandafter\def\csname LT2\endcsname{\color{black}}%
      \expandafter\def\csname LT3\endcsname{\color{black}}%
      \expandafter\def\csname LT4\endcsname{\color{black}}%
      \expandafter\def\csname LT5\endcsname{\color{black}}%
      \expandafter\def\csname LT6\endcsname{\color{black}}%
      \expandafter\def\csname LT7\endcsname{\color{black}}%
      \expandafter\def\csname LT8\endcsname{\color{black}}%
    \fi
  \fi
  \setlength{\unitlength}{0.0500bp}%
  \begin{picture}(7200.00,5040.00)%
    \gplgaddtomacro\gplbacktext{%
      \csname LTb\endcsname%
      \put(990,660){\makebox(0,0)[r]{\strut{} 4.1}}%
      \put(990,1483){\makebox(0,0)[r]{\strut{} 4.2}}%
      \put(990,2306){\makebox(0,0)[r]{\strut{} 4.3}}%
      \put(990,3130){\makebox(0,0)[r]{\strut{} 4.4}}%
      \put(990,3953){\makebox(0,0)[r]{\strut{} 4.5}}%
      \put(990,4776){\makebox(0,0)[r]{\strut{} 4.6}}%
      \put(1594,440){\makebox(0,0){\strut{} 5}}%
      \put(2774,440){\makebox(0,0){\strut{} 10}}%
      \put(3954,440){\makebox(0,0){\strut{} 15}}%
      \put(5134,440){\makebox(0,0){\strut{} 20}}%
      \put(6210,660){\makebox(0,0)[l]{\strut{} 0}}%
      \put(6210,1093){\makebox(0,0)[l]{\strut{} 0.2}}%
      \put(6210,1527){\makebox(0,0)[l]{\strut{} 0.4}}%
      \put(6210,1960){\makebox(0,0)[l]{\strut{} 0.6}}%
      \put(6210,2393){\makebox(0,0)[l]{\strut{} 0.8}}%
      \put(6210,2826){\makebox(0,0)[l]{\strut{} 1}}%
      \put(6210,3260){\makebox(0,0)[l]{\strut{} 1.2}}%
      \put(6210,3693){\makebox(0,0)[l]{\strut{} 1.4}}%
      \put(6210,4126){\makebox(0,0)[l]{\strut{} 1.6}}%
      \put(6210,4559){\makebox(0,0)[l]{\strut{} 1.8}}%
      \put(220,2718){\rotatebox{90}{\makebox(0,0){\strut{}$E$ per $B=4$ Skyrmion}}}%
      \put(6979,2718){\rotatebox{90}{\makebox(0,0){\strut{}$|E_\text{int}-E_\text{int}^{l\leq3}|/|E_\text{int}|$}}}%
      \put(3600,110){\makebox(0,0){\strut{}$a$}}%
    }%
    \gplgaddtomacro\gplfronttext{%
      \csname LTb\endcsname%
      \put(5091,4603){\makebox(0,0)[r]{\strut{}Numerically determined crystal energy}}%
      \csname LTb\endcsname%
      \put(5091,4383){\makebox(0,0)[r]{\strut{}Relative error of asymptotic $E_\text{int}^{l\leq3}$}}%
    }%
    \gplbacktext
    \put(0,0){\includegraphics{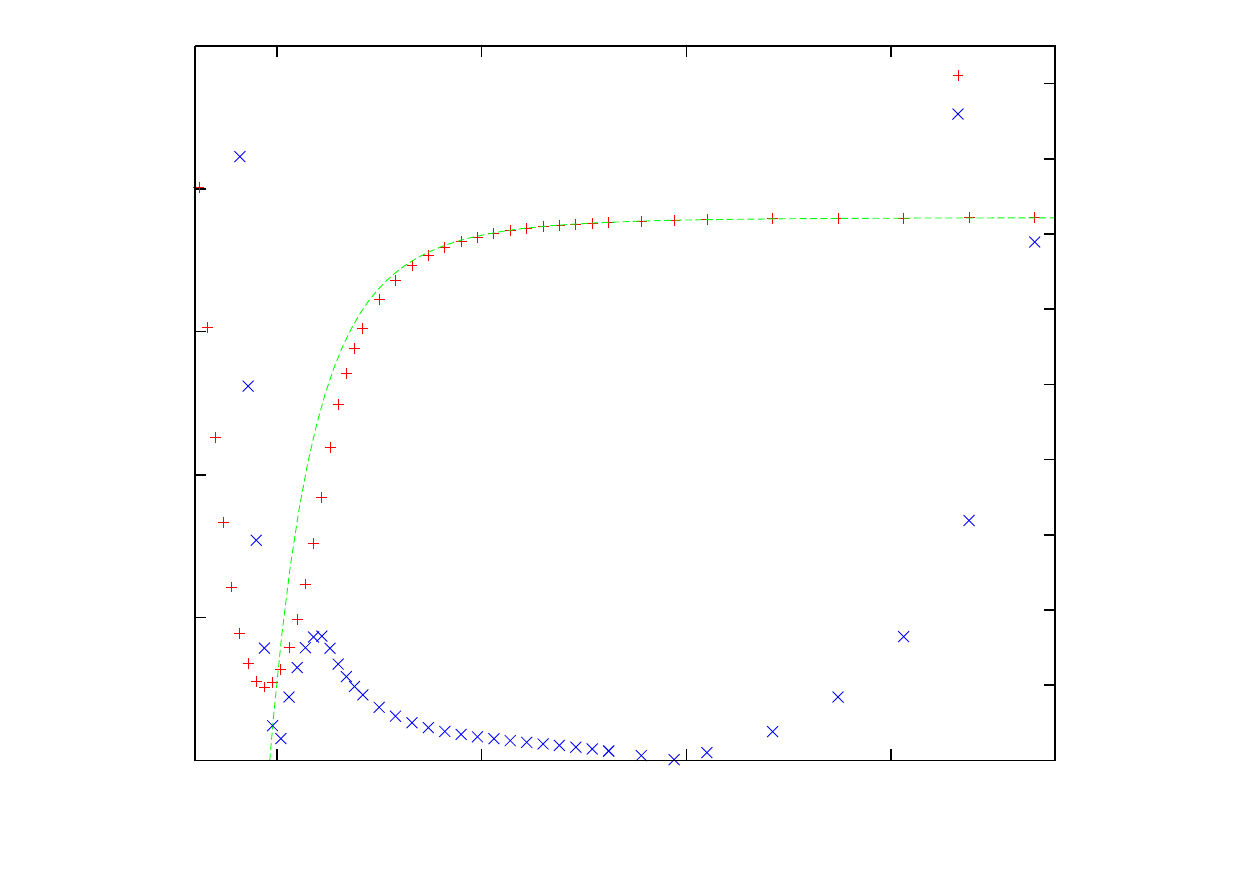}}%
    \gplfronttext
  \end{picture}%
\endgroup